\documentclass[aps,reprint,prb,superscriptaddress]{revtex4-2}
\usepackage{appendix}
\usepackage{xcolor}
\usepackage{amsmath,amssymb,amsthm}
\usepackage{bm}
\usepackage{comment}
\usepackage[normalem]{ulem}
\usepackage{siunitx}
\usepackage{grffile}
\usepackage{algorithm}
\usepackage{algpseudocode}
\newcommand{\rmd}{{\rm d}}

\newcommand{\kB}{k_{\rm B}}

\newcommand{\vF}{v_{\rm F}}
\DeclareMathOperator{\sgn}{sgn}

 \usepackage[
 colorlinks=true,
 citecolor=gray,linkcolor=gray, urlcolor=gray % Any Color You Like
 ]{hyperref}

\graphicspath{{./images/}}

\begin{document}
%
% Use the \preprint command to place your local institutional report
% number in the upper righthand corner of the title page in preprint mode.
% Multiple \preprint commands are allowed.
% Use the 'preprintnumbers' class option to override journal defaults
% to display numbers if necessary
%\preprint{}

%Title of paper
\title{Chirality-dependent spin polarization in metals:\\
linear and quadratic responses}
% repeat the \author .. \affiliation  etc. as needed
% \email, \thanks, \homepage, \altaffiliation all apply to the current
% author. Explanatory text should go in the []'s, actual e-mail
% address or url should go in the {}'s for \email and \homepage.
% Please use the appropriate macro foreach each type of information
% \affiliation command applies to all authors since the last
% \affiliation command. The \affiliation command should follow the
% other information
% \affiliation can be followed by \email, \homepage, \thanks as well.
\author{Kosuke Yoshimi}
\email{yoshimi-kosuke567@g.ecc.u-tokyo.ac.jp}
\affiliation{Department of Physics, Graduate School of Science, The University of Tokyo, 7-3-1 Hongo, Tokyo 113-0033, Japan}
\author{Yusuke Kato}
\email{yusuke@phys.c.u-tokyo.ac.jp}
\affiliation{Department of Physics, Graduate School of Science, The University of Tokyo, 7-3-1 Hongo, Tokyo 113-0033, Japan}
\affiliation{Department of Basic Science, The University of Tokyo, 3-8-1 Komaba, Tokyo 153-8902, Japan}
\affiliation{Quantum Research Center for Chirality, Institute for Molecular Science, Okazaki, Aichi 444-8585, Japan}

\author{Yuta Suzuki}
\affiliation{Department of Physics, Institute of Science Tokyo, 2-12-1 Ookayama, Tokyo 152-8551, Japan}
%\email{suzuki@vortex.c.u-tokyo.ac.jp}
%\altaffiliation[Present address: ]{Department of Physics, Institute of Science Tokyo, 2-12-1 Ookayama, Tokyo 152-8551, Japan}
%\affiliation{Department of Physics, Graduate School of Science, The University of Tokyo, 7-3-1 Hongo, Tokyo 113-0033, Japan}

\author{Shuntaro Sumita}
\affiliation{Department of Basic Science, The University of Tokyo, 3-8-1 Komaba, Tokyo 153-8902, Japan}
\affiliation{RIKEN Center for Emergent Matter Science, Wako, Saitama 351-0198, Japan}

\author{Takuro Sato}
%\email{takurosato@ims.ac.jp}
\affiliation{Research Center of Integrative Molecular System, Institute for Molecular Science, National Institutes of Natural Sciences, Okazaki, Aichi 444-8585, Japan}
\author{Hiroshi M.~Yamamoto}
%\email{yhiroshi@ims.ac.jp}
\affiliation{Research Center of Integrative Molecular System, Institute for Molecular Science, National Institutes of Natural Sciences, Okazaki, Aichi 444-8585, Japan}
\affiliation{Quantum Research Center for Chirality, Institute for Molecular Science, Okazaki, Aichi 444-8585, Japan}
\author{Yoshihiko Togawa}
%\email{ytogawa@omu.ac.jp}
\affiliation{
Department of Physics and Electronics, Osaka Metropolitan University, 1-1 Gakuencho, Sakai, Osaka 599-8531, Japan }
\affiliation{Quantum Research Center for Chirality, Institute for Molecular Science, Okazaki, Aichi 444-8585, Japan}
\author{Hiroaki Kusunose}
%\email{hk@meiji.ac.jp}
\affiliation{Department of Physics, Meiji University, Kawasaki 214-8571, Japan }
\affiliation{Quantum Research Center for Chirality, Institute for Molecular Science, Okazaki, Aichi 444-8585, Japan}
\author{Jun-ichiro Kishine}
%\email{kishine@ouj.ac.jp}
\affiliation{Division of Natural and Environmental Sciences, The Open University of Japan, Chiba 261-8586, Japan }
\affiliation{Quantum Research Center for Chirality, Institute for Molecular Science, Okazaki, Aichi 444-8585, Japan}
\affiliation{Department of Basic Science, The University of Tokyo, 3-8-1 Komaba, Tokyo 153-8902, Japan}

%\homepage[]{Your web page}
%\thanks{}
%Collaboration name if desired (requires use of superscriptaddress
%option in \documentclass). \noaffiliation is required (may also be
%used with the \author command).
%\collaboration can be followed by \email, \homepage, \thanks as well.
%\collaboration{}
%\noaffiliation

\date{\today}

%%%%%%%%%%%%%%%%%%%%%%%%%%%%%%%%%%%%%%%%%%%%%%%%%%%%%%%
\begin{abstract}% <600 characters for PRL.
We study spin polarization induced by locally injected electric currents in a metal whose spin--orbit coupling
 reflects its structural chirality.
We reveal both spin polarization in the bulk in the linear response and antiparallel spin polarization near the interface in the quadratic response to external electric currents, and reproduce the experimentally observed correlation between the chirality of the metal and the direction of spin polarization.
In particular, we elucidate that the sign of the spin polarization in the quadratic response is opposite to that expected from the bulk spin current.
This sign discrepancy originates from spin polarization induced by dipole-like charge distribution appearing in the quadratic response.
%This sign discrepancy originates from the dipole-like charge distribution appearing in the quadratic response.
%Our method is applicable to a wide range of real materials with various types of spin--orbit %coupling.
\end{abstract}
%%%%%%%%%%%%%%%%%%%%%%%%%%%%%%%%%%%%%%%%%%%%%%%%%%%%%%%
\maketitle
%%%%%%%%%%%%%%%%%%%%%%%%%%%%%%%%%%%%%%%%%%%%%%%%%%%%%%%

%%%%%%%%%%%%%%%%%%%%%%%%%%%%%%%%%
\section{Introduction}
\begin{figure}
 \centering
 \includegraphics[width=1.0\linewidth]{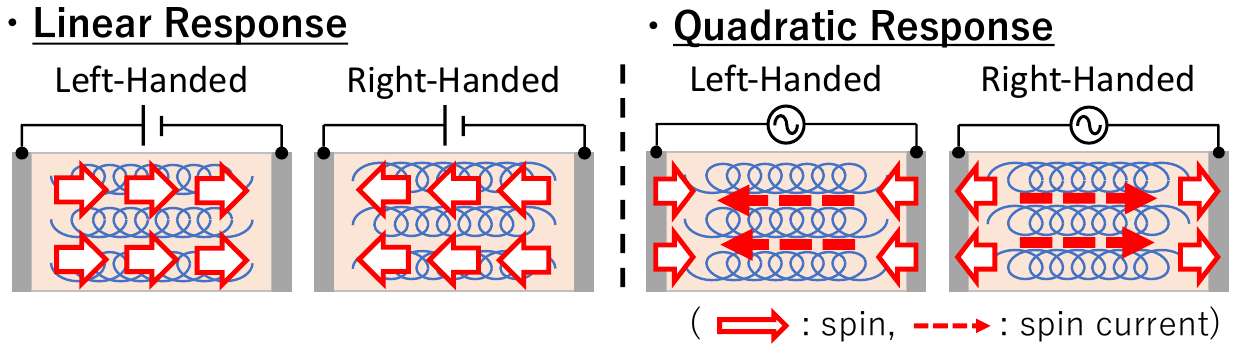}
 \caption{Schematics for the earlier studies on the linear response against the DC driving current~\cite{Inui2020,Nabei2020,Shiota2021,Shishido2021,Shishido2023} and those on the quadratic response to the AC driving current~\cite{nakajima2023,Nakajima_Thesis}.
 They show the chirality-dependent spin polarization, and the antiparallel spin near interfaces in the quadratic response. The dotted arrows in the right panel indicate the flow direction of carriers with rightward polarized spin.}
 \label{fig: introduction}
\end{figure}

Chirality-induced spin selectivity (CISS)~\cite{Naaman2012,Naaman2015,Naaman2018,Naaman2019,Naaman2019apl,Naaman2020-jp,Naaman2020-vz,Waldeck2021,Aiello2022,Bloom2024}, first observed 
in photoelectron transmission through chiral molecules~\cite{Ray1999,Goehler2011}, has drawn significant attention due to large spin polarization at room temperature. 
Following the early studies, related phenomena have been reported in various systems, ranging from chiral molecules to inorganic metals~\cite{Inui2020, Nabei2020,Shiota2021,Shishido2021,Shishido2023} and a superconductor~\cite{nakajima2023,Nakajima_Thesis}. 
Accordingly, the term CISS now refers to the general correlation between electron spin polarization and the chirality of materials ~\cite{Naaman2020-jp,Naaman2020-vz,Waldeck2021,Evers2022,Bloom2024,sala2026,Suda2019,aizawa2023}.
In this paper, we focus on the spin polarization in chiral metals, in the setup analogous to the experiments that demonstrated the CISS in solid state physics in the {\it linear} ~\cite{Inui2020,Nabei2020,Shiota2021,Shishido2021,Shishido2023} and the {\it quadratic} ~\cite{nakajima2023,Nakajima_Thesis} responses to a {\it locally}-injected electric current.  
We develop a microscopic theory that can consistently describe two features observed in these experiments: (i) chirality-dependent spin polarization and (ii) antiparallel spin polarization near interfaces, as illustrated in Fig.~\ref{fig: introduction}.
We also show (iii) that conventional estimates based solely on spin current are not necessarily sufficient to capture the features of the spin polarization, as likewise noted in the earlier studies of the spin Hall effect and spin Nernst effect~\cite{ShitadeTatara2022,Shitade2022}. Our findings demonstrate the importance of a theoretical scheme capable of addressing both spin and charge polarizations near the interface of chiral crystals. 

We analytically address those points (i), (ii), and (iii) by solving the Boltzmann equation that explicitly incorporates the local current injection, together with the Gauss's law to take account of electric polarization near the interfaces.

This paper is organized as follows.
In Sec.~II, we present our model of a chiral system and the Boltzmann equation satisfying charge conservation.
In Sec.~III, we apply our method to a bulk chiral system as a prerequisite.
In Sec.~In IV, we present the calculated results for linear and quadratic spin responses in an interfacial system and discuss the sign of the spin accumulation near the interfaces.
In Appendix~\ref{appendix_ResultsOverview}, we summarize the principal results for clarity and readability.
In Appendix~\ref{appendix_DetailedCalculations}, we summarize the derivation of the linear and quadratic spin responses in the interfacial system and present the spin polarization for several models of local current injection.
\color{black}
Throughout this paper, we set $\hbar = 1$ and take $e > 0$.

%%%%%%%%%%%%%%%%%%%%%%%%%%%%%%%%%
\section{Model}
\begin{figure}
 \includegraphics[pagebox=artbox,width=0.47\textwidth]{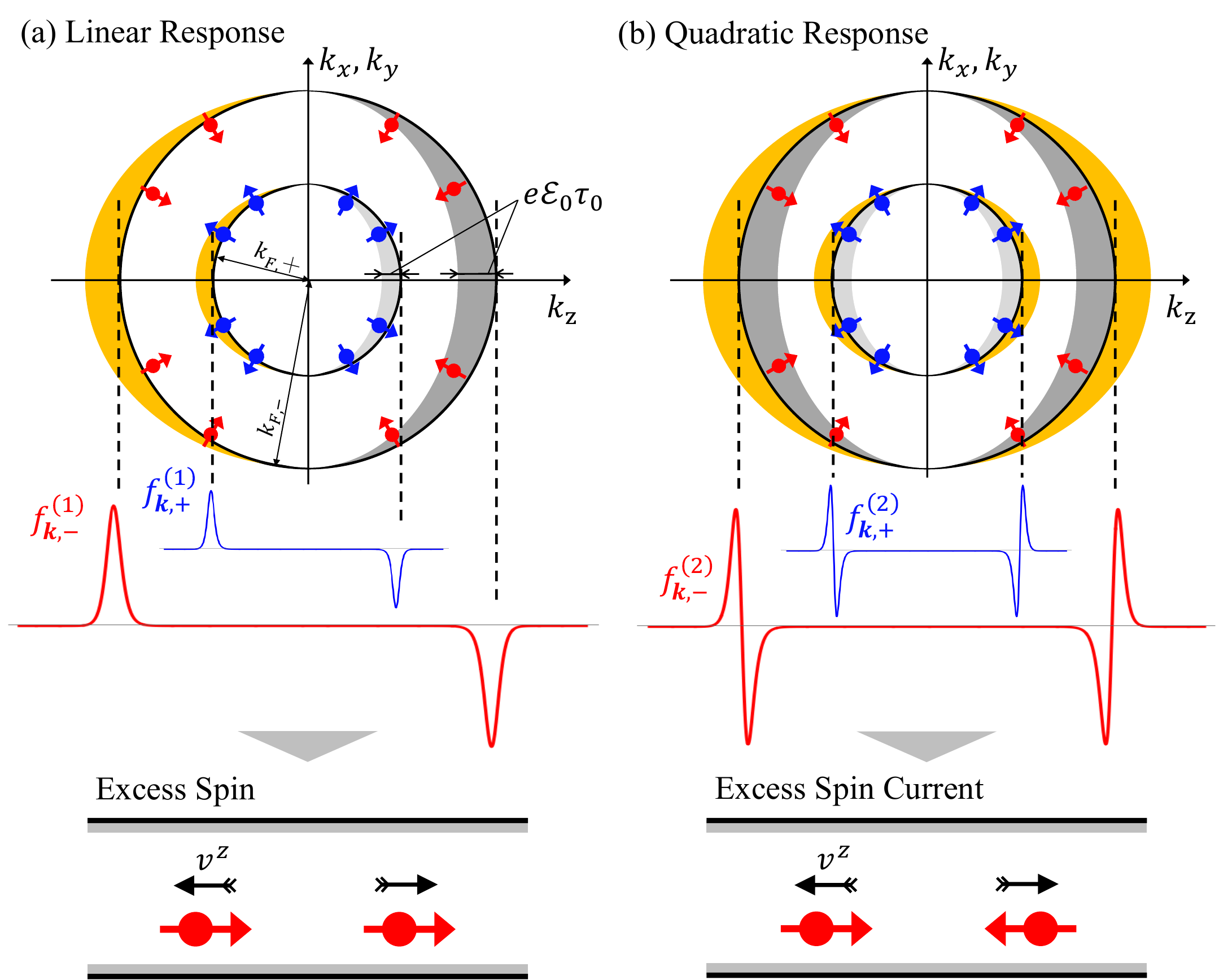}
 \caption{
 Schematics for the deviation of the distribution function around the Fermi surfaces of the spin-splitting bands in (a) linear and (b) quadratic responses, for $q=-e$ and $\alpha > 0$. The shaded regions in orange (gray) denote the excess (deficit) distributions. These excess distributions \textcolor{black}{stem from} differences in the density of states \textcolor{black}{between the spin-splitting band}, which, for instance in the linear response, leads to a net spin density. The arrows represent the direction of spin in each Bloch state. These figures account for the bulk responses shown in Eqs.~\eqref{eq:bulk-response-s1} and \eqref{eq:bulk-response-j2}.}
 \label{fig: momentumspaceschematics}
\end{figure}

As a minimal model, we take a metal with spin–orbit coupling (SOC) characteristic of chiral systems, expressed in an isotropic form. 
Here we use the word ``chiral'' for the 3D system with time-reversal symmetry without inversion centers, mirror planes, roto-inversion axes, or roto-reflection axes~\cite{barron2012,barron2020,Kishine-Israel,kusunose2024emergence}.
The Hamiltonian is given by\cite{tatara2022,yamaguchi2024}
\begin{equation}
 \mathcal{H}=\frac{k^2}{2m}+\alpha \bm{\sigma}\cdot\bm{k},
 \label{eq: Hamiltonian}
\end{equation}
with $k=|\bm{k}|$ and the Pauli matrices \textcolor{black}{ $\bm{\sigma}=(\sigma_x,\sigma_y,\sigma_z)$ }in the spin space.
The second term represents the hedgehog type antisymmetric SOC \cite{Onuki2014,Furukawa2017,Furukawa2021,Shiota2021}, where the sign of $\alpha$ corresponds to the left/right chirality of the system.
The Hamiltonian is diagonalized as
$
\mathcal{H} |\bm{k},\pm\rangle
= \varepsilon_{\pm}(k) |\bm{k},\pm\rangle,
$
where the energy dispersion relation of the spin-splitting band is given by
$
\varepsilon_{\pm}(k)=\frac{k^2}{2m}\pm |\alpha| k.
$
Note that the spin is parallel (antiparallel) to the wave vector in the $+$ band when $\alpha>0$ ($\alpha<0$); see also Fig.~\ref{fig: momentumspaceschematics}.
The Fermi wave vectors for the spin-splitting band for the chemical potential $\mu$ ($>0$) are given by $k_{\rm F,\pm}=m(\mp |\alpha|+\vF)$ with $\vF=\sqrt{\alpha^2 +2\mu/m}$.

Using this model, we discuss linear and quadratic spin responses in the chiral metal to an external DC current density $j_0$ along the $z$ direction.
For the band $\gamma = \pm$, let $f_{\bm{k}\gamma}$ be the distribution function and $\bm{v}_{\bm{k}\gamma} := \partial\varepsilon_\gamma / \partial\bm{k}$ the velocity.
We use the Boltzmann equation in the presence of the nonmagnetic impurities,
\begin{equation}
 \frac{\partial f_{\bm{k}\gamma}}{\partial t} + v_{\bm{k}\gamma}^z \frac{\partial f_{\bm{k}\gamma}}{\partial z} +q E(z) \frac{\partial f_{\bm{k}\gamma}}{\partial k_z}
 = {\rm St}[f_{\bm{k}\gamma}] + I_{\bm{k}\gamma}(z),
 \label{eq: Boltzmann equation}
\end{equation}
where $q\,(=\pm e)$ is the charge of the carrier and $E(z)$ is the electric field, which directs the $z$-direction.
The first term in the RHS is the collision integral,
\begin{equation}
 {\rm St}[f_{\bm{k}\gamma}]=\frac{1}{\Omega}\sum_{\bm{k'}\gamma'}W({\bm{k}\gamma} \to  {\bm{k}'\gamma'})(f_{\bm{k}'\gamma'}-f_{\bm{k}\gamma})\delta (\varepsilon_{\bm{k}\gamma}-\varepsilon_{\bm{k}'\gamma'}),
 \label{eq: St}
\end{equation}
where $\Omega$ denotes the volume of the system and $W({\bm{k}\gamma} \to {\bm{k}'\gamma'})$ the transition probability from the one-particle state $\bm{k}\gamma$ to another $\bm{k}'\gamma'$.
The second term $I_{\bm{k}\gamma}(z)$ represents the source term, which becomes relevant in later discussions on local injection of electric current.

As for the collision term, we assume isotropy of the scattering probability, and replace $W({\bm{k}\gamma} \to {\bm{k}'\gamma'})$ with $\tilde{W}(\varepsilon_{\bm{k}\gamma})$.
Equation~\eqref{eq: St} then reduces to
\begin{equation}
 {\rm St}[f_{\bm{k}\gamma}]
 \to
 {\rm \tilde{St}}[f_{\bm{k}\gamma}]
 =
 \frac{n(\varepsilon_{\bm{k}\gamma})}{N(\varepsilon_{\bm{k}\gamma})\tau(\varepsilon_{\bm{k}\gamma})}-\frac{f_{\bm{k}\gamma}-f^{(0)}(\varepsilon_{\bm{k}\gamma})}{\tau(\varepsilon_{\bm{k}\gamma})},
 \label{eq: St-constant-W}
\end{equation}
with the Fermi distribution function $f^{(0)}(\varepsilon)$.
The first term on the RHS of Eq.~\eqref{eq: St-constant-W} is essential for satisfying the charge conservation law in interface problems, even though it is typically excluded in the conventional relaxation time approximation.
Here $\tau(\varepsilon)$ is a relaxation time defined by $\frac{1}{\tilde{W}(\varepsilon)N(\varepsilon)}$ in terms of the density of one-particle states $N(\varepsilon)=\frac{1}{\Omega}\sum_{\bm{k}\gamma}\delta(\varepsilon-\varepsilon_{\bm{k}\gamma})$.
For simplicity, we replace $\tau(\varepsilon)$ by a constant $\tau_0$.
The symbol $n(\varepsilon_{\bm{k}\gamma})$ denotes
\begin{equation}
 n(\varepsilon)=\frac{1}{\Omega}\sum_{\bm{k}\gamma}[f_{\bm{k}\gamma}-f^{(0)}(\varepsilon)]\delta (\varepsilon_{\bm{k}\gamma}-\varepsilon),
 \label{eq: n-epsilon}
\end{equation}
which represents the excess carrier density stemming from the one-particle states with energy $\varepsilon$.
We define the $i$th order ($i=1,2$) of the spin density, namely spin polarization, and the spin current density with respect to  the external electric current density $j_0$, by
\begin{subequations}
 \begin{align}
  s^{(i)}_z(z,t) &=
  \frac{1}{\Omega}\sum_{\bm{k}\gamma}\Big\langle\bm{k},\gamma\Big|\frac{\sigma_z}{2}\Big|\bm{k},\gamma\Big\rangle f^{(i)}_{\bm{k}\gamma}(z,t),
  \label{eq: s-def} \\
  j^{(i)}_{s;zz}(z,t) &=
  \frac{1}{\Omega}\sum_{\bm{k}\gamma}\Big\langle\bm{k},\gamma\Big|\frac{\sigma_z}{2}v_{\bm{k}\gamma}^z \Big|\bm{k},\gamma\Big\rangle f^{(i)}_{\bm{k}\gamma}(z,t).
 \end{align}
\end{subequations}
Here $f^{(i)}_{\bm{k}\gamma}(z,t)$ denotes the distribution function of the $i$th order \textcolor{black}{ with respect to $j_0$}.
In the following, we calculate these spin responses at low temperatures so that $\kB T$ is much smaller than the chemical potential $\mu$.

%%%%%%%%%%%%%%%%%%%%%%%%%%%%%%%%%

\section{Prerequisite: Bulk responses to the uniform electric current}
We first discuss the linear and quadratic responses to a uniform DC electric current without boundaries or the source term $I_{\bm{k}\gamma}$, as a prerequisite to addressing the effect of boundaries or local input discussed below.
In this case, the electric field $E(z)$ is given by a constant $\mathcal{E}_0 := j_0 / \sigma_0$ with the conductivity $\sigma_0= q^2\tau_0\sum_{\bm{k},\gamma}(v^{z}_{\bm{k},\gamma})^2 \delta(\varepsilon_{\bm{k},\gamma}-\mu)/\Omega$.
% Under a static electric field $\bm{E}=\mathcal{E}_0 \bm{e}_z$, with the unit vector $\bm{e}_z$ along the $z$-axis,
The excess charge $q n(\varepsilon)$ is zero everywhere and thus the scattering integral in Eq.~\eqref{eq: St-constant-W} reduces to that of the relaxation time approximation.
We obtain
\begin{subequations}
 \label{eq:bulk-response}
 \begin{align}
  & s^{(1)}_z=-\frac13\frac{q \mathcal{E}_0 \tau_0}{\pi^2}m^2\alpha \vF, \quad
  j^{(1)}_{s;zz} =0,
  \label{eq:bulk-response-s1} \\
  & s^{(2)}_z=0, \quad
  j^{(2)}_{s;zz}=-\frac{1}{3}\frac{(q \mathcal{E}_0 \tau_0)^2}{\pi^2}m\alpha \vF,
  \label{eq:bulk-response-j2}
 \end{align}
\end{subequations}
by a calculation similar to that in Ref.~\cite{Hamamoto2017}.
Figure~\ref{fig: momentumspaceschematics} qualitatively accounts for the linear and quadratic responses shown in Eqs.~\eqref{eq:bulk-response-s1} and \eqref{eq:bulk-response-j2}.
In the linear response, the nonequilibrium part of the distribution function is antisymmetric with respect to $k_z$ as shown in Fig.~\ref{fig: momentumspaceschematics}(a). This distribution function yields nonzero spin polarization without the spin current (Eq.~\eqref{eq:bulk-response-s1} ).
In contrast, in the quadratic response, the nonequilibrium part of the distribution function is symmetric with respect to $k_z$ [Fig.~\ref{fig: momentumspaceschematics}(b)], which yields the spin current without spin polarization.
\textcolor{black}{From the symmetry-based consideration, nonzero $s_z^{(1)}$ and $j_{s;zz}^{(2)}$ require the breaking of mirror symmetry $z\rightarrow -z$.}

Spin polarization shown in Eq.~\eqref{eq:bulk-response-s1} is referred to as current-induced magnetization, Rashba-Edelstein effect, or inverse spin-galvanic effect~\cite{Vasko1979a,Levitov1985,Aronov1989,Edelstein1990,Kato2004a,Silov2004,Ganichev2006,Furukawa2017,Furukawa2021,SuzukiKato2023,Jagoda2023,barts2025}.
Expressions for spin current in the quadratic response in different chiral models and related models have been given in earlier studies~\cite{Hamamoto2017,He2019,Hirakida2022,Oiwa2022PRL,Yao2024}.
The proportionality to $\alpha$ in Eq.~\eqref{eq:bulk-response} confirms that the direction of the spin and spin current is determined by the chirality of the system.

\begin{figure*}
 \includegraphics[pagebox=artbox,width=\textwidth]{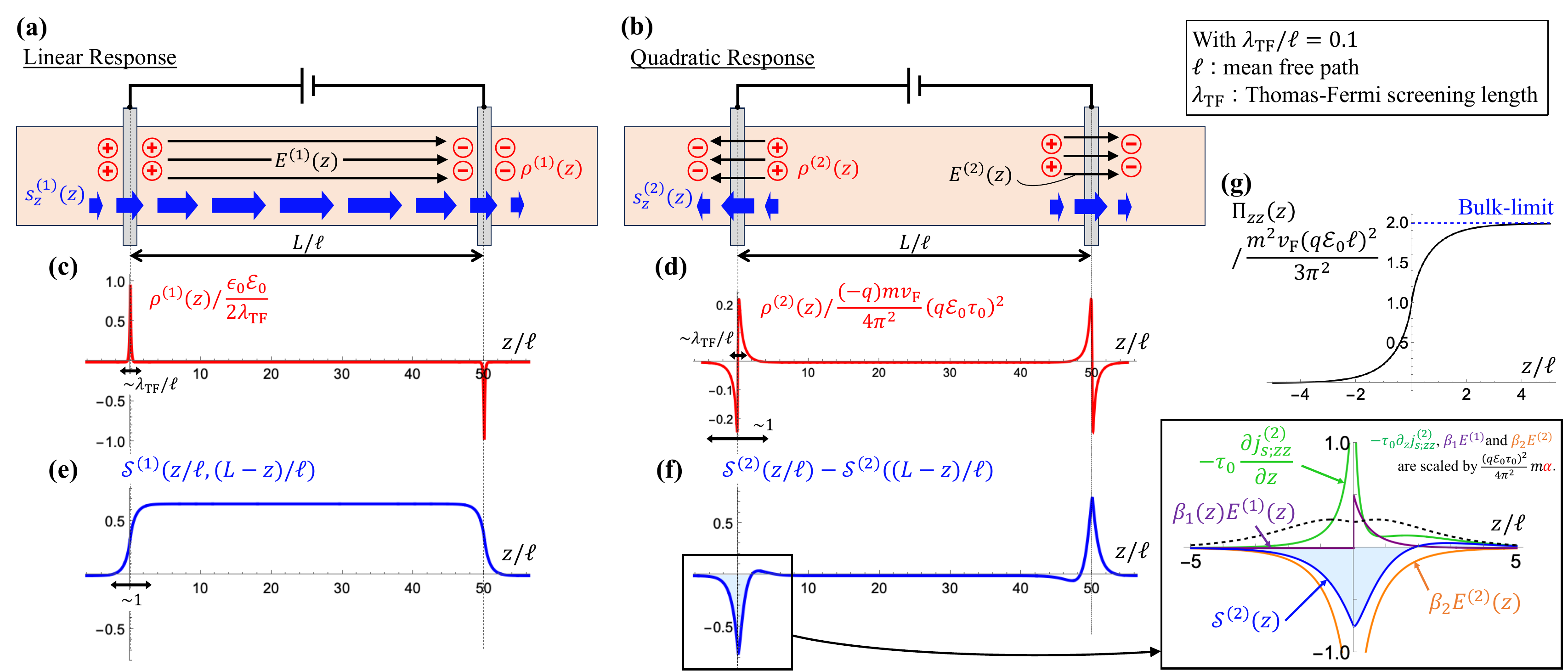}
 \caption{
 (a, b) Schematics of the setup to measure the linear and quadratic responses to the locally uniform electric field.
 Red symbols and blue arrows roughly illustrate the spatial profiles of charge and spin, respectively, for $q=-e$ and $\alpha > 0$.
 (c, d) The spatial dependence of the excess charge density in the linear and quadratic responses, $\rho^{(1)}$ and $\rho^{(2)}$.
 (e, f) Spatial dependence of spin polarization in the linear and quadratic responses.
 The inset shows the enlarged view of $\mathcal{S}^{(2)}(z)$ (blue line).
 The green, purple and orange solid lines show the three components $(-\tau_0)\partial j_{s;zz}^{(2)}/\partial z$, $\beta_1(z)E^{(1)}(z)$ and $\beta_2E^{(2)}(z)$ scaled by $(q\mathcal{E}_0\tau_0)^2m\alpha/4\pi^2$, respectively, whereas the black dashed line shows $s_z^{(2)}(z)$ obtained by the conventional relaxation time approximation neglecting the contribution of the excess electron density in Eq.~\eqref{eq: St-constant-W}.
 For clarity, we set $\lambda_{\mathrm{TF}} / \ell = 0.1$ and $L / \ell = 50$ in panels (c--f).
 (g) The monotonic behavior of $\Pi_{zz}(z)$ near the interface $z = 0$ in the limit of $\lambda_{\mathrm{TF}}/\ell \to 0$.
 }
 \label{fig: localfield}
\end{figure*}

%%%%%%%%%%%%%%%%%%%%%%%%%%%%%%%%%
\section{Spin distribution under a local electric current}
The bulk system of the previous section is now modified by adding interfaces for current flow.
In the following, we consider the linear and quadratic responses to a \textit{local} injection of electric current density $j_0$ at $z=0$ and an extraction at $z=L$ in the system occupying $z \in (-\infty,\infty)$ (see Figs.~\ref{fig: localfield}(a) and \ref{fig: localfield}(b)).
We assume that the current path length $L$ is much longer than the mean free path $\ell:=v_\mathrm{F}\tau_0$, so that sufficiently far from the interfaces, Eqs.~\eqref{eq:bulk-response-s1} and \eqref{eq:bulk-response-j2} are expected to hold.
In the presence of the current injection and extraction, the source term in Eq.~\eqref{eq: Boltzmann equation} is given by
\begin{equation}
 I_{\bm{k}\gamma}(z)
 = J_{\bm{k}\gamma}\left[\delta(z)-\delta(z-L)\right]
 \label{eq: source term}
\end{equation}
with $J_{\bm{k}\gamma}$ satisfying
\begin{equation}
 \frac{1}{\Omega}\sum_{\bm{k}\gamma}qJ_{\bm{k}\gamma}=j_0.
 \label{eq: source-sum-rule}
\end{equation}
This source term was used in the theory of hot-electron transport through thin dielectric films in Ref.~\cite{Bernasconi1988}.
By multiplying Eq.~\eqref{eq: Boltzmann equation} by $q/\Omega$ and summing over $\bm{k}$ and $\gamma$, we obtain the charge conservation law,
\begin{equation}
 \frac{\partial \rho}{\partial t} +\frac{\partial j_e(z)}{\partial z}=j_0 \left[\delta(z)-\delta(z-L)\right],
 \label{eq: charge-conservation}
\end{equation}
with $\rho=\frac{1}{\Omega}\sum_{\bm{k}\gamma}qf_{\bm{k}\gamma}$ and $j_e=\frac{1}{\Omega}\sum_{\bm{k}\gamma}qv^{z}_{\bm{k}\gamma}f_{\bm{k}\gamma}$.
The explicit expression for $J_{\bm{k}\gamma}$ depends on the property of the interface between the chiral metal and the lead of the current source.
We here take the simplest form,
\begin{equation}
 J_{\bm{k}\gamma} = -\frac{3j_0}{2q N_\gamma} \left(\frac{v_{\bm{k}\gamma}^z}{\vF}\right)^2
 \frac{\partial f^{(0)}(\varepsilon_{\bm{k}\gamma})}{\partial \varepsilon_{\bm{k}\gamma}}
 \label{eq: Jkgamma_1}
\end{equation}
with $N_\gamma = \frac{1}{\Omega} \sum_{\bm{k}}\delta(\varepsilon_{\bm{k}\gamma}-\varepsilon)$, the lowest-order choice in $\cos{\theta}:=k_z/k_{\mathrm{F}}$.
This form satisfies Eq.~\eqref{eq: source-sum-rule} and the continuity of the spin density and spin current at $z = 0, \, L$.
We confirm that the final results are not qualitatively changed for other forms of $J_{\bm{k}\gamma}$, except for the breakdown of local Ohm's law (see Appendix~\ref{appendix_DetailedCalculations} for details).

% In the present set-up shown in Fig.~\ref{fig: localfield}, the electric field $E$ and excess electron density $n(\varepsilon)$ varies spatially near $z=0$ and $L$.
The Boltzmann equation \eqref{eq: Boltzmann equation} with Eqs.~\eqref{eq: St-constant-W} and \eqref{eq: source term}, and Gauss's law $\partial E/\partial z=\rho/\epsilon_0=q \int d \varepsilon n(\varepsilon)/\epsilon_0$ with the electric constant (vacuum permittivity) $\epsilon_0$ form a closed set of equations to determine $f_{\bm{k}\gamma}$, $n(\varepsilon)$ and $E$.
Imposing the boundary conditions $f_{\bm{k}\gamma} \to f^{(0)}(\varepsilon_{\bm{k}\gamma})$ and $E \to 0$ in the limits $z \to \pm\infty$, and assuming that $n(\varepsilon)$ is proportional to $-\frac{\partial f^{(0)}}{\partial \varepsilon}\simeq \delta(\varepsilon - \mu)$, we analytically solve the equations up to the second order in $j_0$.

In the linear response, the excess charge density $\rho^{(1)}(z)$ is illustrated in Fig.~\ref{fig: localfield}(c), which shows peaks with the width of the Thomas--Fermi screening length $\lambda_{\rm TF}=\sqrt{\epsilon_0/(q^2 N_0)}$ [$N_0 := N(\mu)$] at the two interfaces.
In typical metals, $\lambda_{\rm TF}\ll\ell$.
Using Eq.~\eqref{eq: s-def}, the spin polarization is given by
\begin{equation}
 s^{(1)}_z(z)=-\frac{q \mathcal{E}_0 \tau_0}{2\pi^2}m^2\alpha \vF\mathcal{S}^{(1)}(z/\ell,(L-z)/\ell),
 \label{eq: s-one}
\end{equation}
where $\mathcal{S}^{(1)}$ is a dimensionless function shown in Fig.~\ref{fig: localfield}(e) (see Appendix~\ref{appendix_ResultsOverview} for its explicit expression).
This result is consistent with the bulk spin polarization of Eq.~\eqref{eq:bulk-response-s1} and the experiments \cite{Inui2020,Nabei2020,Shiota2021,Shishido2021,Shishido2023} due to its proportionality to $\alpha$.

Let us move on to the quadratic response.
We determine charge distributions so that the electric current $j_e^{(2)}(z)$ becomes zero throughout the system~\cite{Note1}.
Figure~\ref{fig: localfield}(d) shows the excess charge density $\rho^{(2)}(z)$, which indicates \textit{dipole}-like charge distribution at the current injection points $z = 0, \, L$.
The electric field $E^{(2)}(z)$ stemming from the electric dipoles (See Fig.~\ref{fig: localfield}(b)(d)) plays an essential role in discussing the spin polarization near the interfaces (See Fig.~\ref{fig: localfield}(b)(f)), as we will argue in the following paragraph.
In the limit of $\lambda_{\rm TF}/\ell \to 0$, the spin polarization is given by
\begin{align}
 s^{(2)}_z(z)=\frac{(q \mathcal{E}_0 \tau_0)^2}{4\pi^2}m\alpha
 \left[\mathcal{S}^{(2)}(z/\ell)-\mathcal{S}^{(2)}((L-z)/\ell)\right],
 \label{eq: s-two}
\end{align}
where $\mathcal{S}^{(2)}(\xi)$ is the dimensionless function shown in Fig.~\ref{fig: localfield}(f) (see Appendix~\ref{appendix_ResultsOverview} for its explicit expression).  Consistent with \eqref{eq:bulk-response-j2}, Eq.~\eqref {eq: s-two}  vanishes in the bulk region ($\ell \ll  z$ and $L-\ell \gg z$), resulting from the cancellation between the two terms in the parentheses in  \eqref{eq: s-two}.   
The spin polarizations near the two interfaces are antiparallel to each other; their sign depends on the chirality parameter $\alpha$, analogous to the experiments \cite{Banerjee-Ghosh2018, nakajima2023}.
For metals, the relative ratio $|s_z^{(2)}(z)/s_z^{(1)}(z)|\sim e \mathcal{E}_0 \tau_0/k_{\rm F}$ is much smaller than unity.
Thus, for sufficiently small driving DC currents, the spin density has the same sign at every point in the system, as observed in Refs.~\cite{Inui2020,Shiota2021,Shishido2021}.
Meanwhile, our results also hold under AC driving at frequencies $\omega$ lower than the relaxation rate $1/\tau_0$.
%In this regime, the lock-in detector allows measurement of the quadratic response.
In this regime, the linear response averages out over time, and the quadratic response becomes dominant.  \textcolor{black}{In the quasi-static limit $\omega \tau_0\ll 1$, the instantaneous response follows the same nonlinear DC expansion, so that the AC response is governed by the same first- and second-order coefficients.}

We demonstrate that the spin polarization around $z=0$ in the quadratic response cannot be explained by a conventional argument solely considering the spin accumulation due to the spin current flowing into the region around $z=0$. 
Along with the conventional argument, %The spin current in the bulk region given by Eq.~\eqref{eq:bulk-response-j2} 
one  would expect spin polarization around $z=0$ with the sign {\it opposite} to that of $j^{(2)}_{s;zz}$ in the bulk region (see Eq.~\eqref{eq:bulk-response-j2}). However, our calculation shows that  the spin polarization around  $z = 0$, which is the area of the blue shaded region in Fig.~\ref{fig: localfield}(f), is $-(4/15\pi^2)(q \mathcal{E}_0 \tau_0)^2 m\alpha \ell$. Namely, the spin polarization around $z=0$ has the {\it same} sign as $j^{(2)}_{s;zz}$ in Eq.~\eqref{eq:bulk-response-j2}.
To clarify the origin of spin  polarization around $z=0$, we utilize the spin balance equation,
\begin{align}
    \frac{\partial j_{s;zz}^{(2)}}{\partial z}-\frac{\beta_1(z)E^{(1)}(z)}{\tau_0}-\frac{\beta_2E^{(2)}(z)}{\tau_0}=-\frac{s_z^{(2)}(z)}{\tau_0},
    \label{eq: spin_transport_equation}
\end{align}
with $\beta_{i}:=\frac{-\tau_0}{2\Omega}\sum_{\bm{k}\gamma}\gamma\cos\theta q\partial_{k_z} f_{\bm{k}\gamma}^{(2-i)}$.  Equation~ \eqref{eq: spin_transport_equation} can be obtained by multiplying Eq.~\eqref{eq: Boltzmann equation} by $\langle\bm{k},\gamma|\sigma_z|\bm{k},\gamma\rangle/2\Omega$ and summing over $\bm{k}$ and $\gamma$ (see Appendix~\ref{appendix_ResultsOverview} for their explicit expressions). 
We show the three contributions $(-\tau_0)\partial j_{s;zz}^{(2)}/\partial z$, $\beta_1(z)E^{(1)}(z)$ and $\beta_2E^{(2)}(z)$ to $s_z^{(2)}(z)$ as the green, purple and orange lines, respectively, in the inset of Fig.~\ref{fig: localfield}(f).
These plots indicate that the negative spin polarization $s_z^{(2)}(z)$ originates from $\beta_2 E^{(2)}(z)$, which was not discussed in previous studies of bulk spin current response~\footnote{Note that this discrepancy does not arise without implementing the charge-conserving modifications to the collision term of the Boltzmann equation [see dashed curve in the inset of Fig.~\ref{fig: localfield}(f)].}.
We note that the coefficient  $\beta_2$ is the same as that of spin density to the electric field in the bulk Edelstein effect under linear response (see Eq.~\eqref{eq:bulk-response-s1}). We thus regard $\beta_2E^{(2)}(z)$ as the spin polarization via Edelstein effect induced by the local electric field $E^{(2)}(z)$ (See Fig.~\ref{fig: localfield}(b)(f)). 
We argue that the sign of $E^{(2)}$ near the interfaces is determined by the sign of charge $q$ of the carrier, on the basis of the force balance relation.
Taking $\sum_{\bm{k}\gamma}mv_{\bm{k}\gamma}^z/\Omega$ of Eq.~\eqref{eq: Boltzmann equation} and integrating it by parts, we obtain the force balance relation  %current $j_e^{(2)}(z)$ is expressed as
\begin{align}
0
 &= -\frac{j_m^{(2)}}{\tau_0}+\frac{q}{\Omega}\sum_{\bm{k}\gamma}\frac{\partial   
  m v_{\bm{k}\gamma}^z}{\partial k_z}(f_{\bm{k}\gamma}^{(1)} E^{(1)}+f_{\bm{k}\gamma}^{(0)} E^{(2)})-\frac{\partial  \Pi_{zz}}{\partial z}\notag \\
 &\sim -\frac{j_m^{(2)}}{\tau_0}+\rho^{(1)}(z)E^{(1)}(z)+\rho^{(0)} E^{(2)}(z)-\frac{\partial\Pi_{zz}}{\partial z}.
 \label{eq: j_e^2(z)}
\end{align}
Here $j_m^{(2)}(=\frac{m j_e^{(2)}}{q})$ denotes the mass current density, which is nothing but the momentum density. In \eqref{eq: j_e^2(z)}, we also introduce the notations $\rho^{(0,1)}=\frac{q}{\Omega}\sum_{\bm{k}\gamma}f^{(0,1)}_{\bm{k}\gamma}$ 
and the momentum flux tensor $\Pi_{zz}(z)$ as
\begin{equation}
 \Pi_{zz}(z) := \frac{1}{\Omega}\sum_{\bm{k}\gamma} m \left(v_{\bm{k}\gamma}^z\right)^2f_{\bm{k}\gamma}^{(2)}(z).
 \label{eq: Pizz}
\end{equation}
We use the approximation $\partial_{k_z} m v_{\bm{k}\gamma}^z \sim 1$ in the last line in \eqref{eq: j_e^2(z)}.
Equation \eqref{eq: j_e^2(z)} represents the balance in this steady state among the viscous force $-\frac{j_m^{(2)}}{\tau_0}$, the Lorentz forces $\rho^{(0)} E^{(2)}(z)+\rho^{(1)}(z)E^{(1)}(z)$, and the hydrodynamic pressure $-\frac{\partial\Pi_{zz}}{\partial z}$.  
In the quadratic response, the mass flow is zero, $j_m^{(2)}=0$. 
Near the current source $z=0$, the dominant contribution in $\rho^{(1)}(>0)$ and $E^{(1)}(>0)$ is localized within the range of charge screening length $\lambda_{\rm TF}$ and thus it is negligible in  \eqref{eq: j_e^2(z)} and $\rho^{(0)} E^{(2)}(z)\sim \frac{\partial\Pi_{zz}}{\partial z}$ for $z\sim O(\ell)$.   
The spatial dependence of  $\Pi_{zz}(z)$ near $z = 0$ is shown in Fig.~\ref{fig: localfield}(g).
We can confirm $\Pi_{zz}(z)>0$ in the current path in the bulk region ($\ell \ll z$ and $L-z\gg \ell$), as 
$\Pi_{zz}(z)/\tau_0=-q\mathcal{E}_0 m\langle(v^z)^2\partial f^{(1)}/\partial k_z\rangle\sim q\mathcal{E}_0 m \langle\partial(v^z)^2/\partial k_z\cdot f^{(1)}\rangle\sim 2\mathcal{E}_0j_0 >0$, which represents the Joule heat and is positive. In the equilibrium state at $z\to -\infty$, $\Pi_{zz}(z)$ is zero.  %$\Pi_{zz}(z)>0$ in the current path (i.e., $0<z<L$) due to the increment of the kinetic energy density. We can confirm that 
%$\Pi_{zz}(z\rightarrow+\infty)/m\tau\sim\mathcal{E}_0J_0[\mathrm{Joule~Heat}]$]
Assuming the monotonicity near $z = 0$,  we find that $-\partial\Pi_{zz}/\partial z<0$, which yields a hydrodynamic pressure to push out the carriers from the current path ($0<z<L$ ) towards the outer region $(z<0)$. 
The Lorentz force  $\rho^{(0)}E^{(2)}(z)$ is thus positive near $z = 0$ to balance with $-\partial_z\Pi_{zz}(z)<0$.
When $q=-e<0$,  $\rho^{(0)}<0$ and thus $E^{(2)}<0$,  and electric dipoles direct inward, as we can infer from the profile of $\rho^{(2)}$ in Fig.~\ref{fig: localfield} (b)(d) and the Gauss's law.  
When the carrier is a hole,  $\rho^{(0)}>0$ and thus $E^{(2)}(z)>0$,  and the electric dipoles direct outward. This dipole-like charge distribution is a general feature not limited to chiral metals since the leading term of $\rho^{(2)}(z)$ is independent of $\alpha$, 

%%%%%%%%%%%%%%%%%%%%%%%%%%%%%%%%%
\section{Summary and discussions}
We addressed the linear and quadratic responses of spin polarization in a 3D isotropic chiral metal under uniform or local DC electric currents.
\color{black}
The key result of this work is that, in mirror-symmetry–broken (chiral) metals, the symmetry-allowed coupling between  $s_z$  and the $z$-component of the electric field leads to a chirality-controlled sign rule for the quadratic spin polarization near interfaces.
While related nonlinear spin responses may arise in other spin–orbit–coupled systems, the specific out-of-plane component $s_z$ considered here is symmetry-forbidden in mirror-symmetric systems, and therefore constitutes a distinct signature of chiral conductors.
The present framework may be extended to other symmetry classes where different spin–electric couplings are allowed, which will be explored in future studies.
We also clarified that, by using the Boltzmann equation satisfying the charge conservation law, \color{black}
the sign of spin polarization at the interfaces in the quadratic response is opposite to that expected from the bulk spin current.
This discrepancy originates from the dipole-like charge distribution. Our results also imply similar responses under  the AC current injection when the frequency is much lower than $1/\tau_0$. 
\begin{comment}
    Our method is readily applicable to other systems, such as those with Hamiltonian including anisotropic spin--orbit coupling.
\end{comment}

To experimentally verify the spin polarization in the quadratic response, a long (or extended) spin relaxation length is required, which is usually found in semimetals and materials with anisotropic SOC.
Our method is applicable to these systems as well, even though this paper focuses on metals with shorter relaxation lengths.

We also remark on an additional length scale associated with an unresolved issue; 
in Refs.~\cite{Inui2020,Nabei2020,Shiota2021,Shishido2021,Shishido2023,nakajima2023,Nakajima_Thesis}, spin polarization was observed hundreds of microns or millimeters away from the local input of charge current, which was dubbed the nonlocal effect.
The mean free path and the spin-diffusion length are far shorter than the length scale for the experimentally observed nonlocal effect.
Thus, the mechanism of nonlocal effects in chiral metals and superconductors remains an open issue to be explored in future studies.

Lastly, we remark on a potential application of the quadratic response in the CISS.
There is growing interest in realizing electromotive forces in inversion-symmetry-broken materials without external bias---for example by utilizing photoinduced shift currents~\cite{Koch1976,Young2012} or local thermal fluctuations instead of macroscopic temperature gradients~\cite{Arisawa2024}---motivated by prospects for diversifying power sources and utilizing waste heat.
In light of these trends, future studies on the quadratic response in the CISS may explore its application to energy harvesting by exploiting spin polarization induced by nonequilibrium fluctuations in chiral materials.
The large spin polarization in the CISS at room temperature can be advantageous for applications, particularly in terms of energy conversion efficiency. Furthermore, by leveraging novel properties such as chiral phonons, it may become feasible to utilize materials not traditionally considered suitable for energy conversion, such as insulators~\cite{ohe2024}.

\section*{Acknowledgments}
We thank H.~Watanabe for his informative comments on the linear and nonlinear responses of chiral systems. K.Y. thanks T. Kimura, T. Kawamura, and Y. Maruyama for the helpful comments. This work was supported by JSPS KAKENHI Grants No.~JP20K03855, No.~JP21H01032, No.~JP22J12348, No.~JP23H00291, No. JP23H00091, No. JP23K03288, No. JP23K20825, No.~JP24KJ1036, No. JP24K01331, No. JP25H02113, PRESTO from JST (Grant Number JPMJPR2356), by Joint Research by the Institute for Molecular Science (IMS program No. 23IMS1101), by the grant of OML Project by the National Institutes of Natural Sciences (NINS program No, OML012301), by World-leading Innovative Graduate Study Program for Materials Research, Information, and Technology (MERIT-WINGS) of the University of Tokyo, and by JST SPRING, Grant Number JPMJSP2108.

%%%%%%%%%%%%%%%%%%%%%%%%%%%%%%%%%%%%%%%%%%%%%%%%%%%%%%%
%\clearpage
\appendix
\onecolumngrid
%%%%%%%%%%%%%%%%%%%%%%%%%%%%%%%%%
\section{\label{appendix_ResultsOverview} Results Overview}
This Appendix A summarizes the principal results for clarity and readability.
In section 1, we derive the results for the case where $J_{\bm{k}\gamma}$ satisfies Eq.~\eqref{eq: Jkgamma_1}.
Section 2 presents the more general forms of $J_{\bm{k}\gamma}$.

\subsection{Analytical expressions for physical quantities}
We show the analytical expressions for the charge density, electric field, and spin polarization, which are obtained by solving the Boltzmann equation~\eqref{eq: Boltzmann equation} with Eqs.~\eqref{eq: source term} and \eqref{eq: Jkgamma_1}, for linear and quadratic responses to the local current injection $j_0$.

Let us begin with the linear response.
The excess charge density and the electric field are written as
\begin{align}
 \rho^{(1)}(z) &= \frac{\epsilon_0 \mathcal{E}_0}{2 \lambda_{\rm TF}}\left(e^{-|z|/\lambda_{\rm TF}}-e^{-|L-z|/\lambda_{\rm TF}}\right), \displaybreak[2] \\
 E^{(1)}(z) &=
 \begin{cases}
  \frac{\mathcal{E}_0}{2}e^{z/\lambda_{\rm TF}} & z < 0, \\
  \mathcal{E}_0-\frac{\mathcal{E}_0}{2}\left(e^{-z/\lambda_{\rm TF}}+e^{(z-L)/\lambda_{\rm TF}}\right) & 0 < z < L, \\
  \frac{\mathcal{E}_0}{2}e^{(L-z)/\lambda_{\rm TF}} & L < z,
 \end{cases}
\end{align}
where $\mathcal{E}_0 = j_0 / \sigma_0$.
The spin polarization is given by Eq.~\eqref{eq: s-one} with a dimensionless function $\mathcal{S}^{(1)}$ defined by
\begin{align}
 \mathcal{S}^{(1)}(\xi,\xi')
 := -\left[\tilde{F}_4(\xi)+\tilde{F}_4(\xi')\right] +
 \begin{cases}
  2/3 & \xi > 0 \ \text{and} \ \xi' > 0, \\
  0 & \text{otherwise},
 \end{cases}
\end{align}
where we define special functions $F_i(\xi):=\int_1^\infty \frac{\rmd u}{u^i}\exp(-u \xi)$ for $\xi>0$ and $\tilde{F}_i(\xi):=(\xi/|\xi|)^{i+1} F_i(|\xi|)$ for $\xi\in (-\infty,\infty)$.

In the quadratic response, the charge density and the electric field are given by
\begin{align}
 \rho^{(2)}(z) &= q\frac{m v_{\mathrm{F}} (q\mathcal{E}_0 \tau_0)^2}{4\pi^3} [(R_1 * R_2)(z / \ell) + (R_1 * R_2)((L-z) / \ell)], \\
 E^{(2)}(z) &= \frac{q \ell}{\epsilon_0} \frac{m v_{\mathrm{F}} (q \mathcal{E}_0 \tau_0)^2}{4\pi^3} [(R_1' * R_2)(z / \ell) - (R_1' * R_2)((L-z) / \ell)],
\end{align}
where $[A * B] := \int_{-\infty}^{\infty} d\xi' \, A(\xi - \xi') B(\xi')$ represents a convolution, and the dimensionless functions $R_1$, $R_1'$ and $R_2$ are defined by
\begin{align}
 & R_1(\xi) =
 \begin{cases}
   \sqrt{\frac{\pi}{2}} \frac{3}{2} e^{\ell \xi / \lambda_{\mathrm{TF}}} & \xi < 0, \\
  - \sqrt{\frac{\pi}{2}} \left(\frac{1}{2} + \frac{\ell \xi}{\lambda_{\mathrm{TF}}}\right) e^{-\ell \xi / \lambda_{\mathrm{TF}}} & \xi > 0,
 \end{cases} \\
 & R_1'(\xi) =
 \begin{cases}
  \sqrt{\frac{\pi}{2}} \frac{3}{2} 
  %\frac{\ell}{\lambda_{\mathrm{TF}}} 
  \frac{\lambda_{\mathrm{TF}}}{\ell}e^{\ell \xi / \lambda_{\mathrm{TF}}} & \xi < 0, \\
  \sqrt{\frac{\pi}{2}} 
  %\left(\frac{3}{2} \frac{\ell}{\lambda_{\mathrm{TF}}} + \xi\right) 
  \left(\frac{3}{2} \frac{\lambda_{\mathrm{TF}}}{\ell} + \xi\right) e^{-\ell \xi / \lambda_{\mathrm{TF}}} & \xi > 0,
 \end{cases} \\
 & R_2(\xi) = - \sqrt{\frac{\pi}{2}} \int_{1}^{\infty} dx \frac{x^3 e^{-x |\xi|}}{(1 - x^2) \left\{ \left[x + \frac{1}{2} \ln\left(\frac{x-1}{x+1}\right)\right]^2 + \frac{\pi^2}{4} \right\}}.
\end{align}
The quadratic spin polarization is given by Eq.~\eqref{eq: s-two} with a dimensionless function $\mathcal{S}^{(2)}(\xi) = \mathcal{J}^{(2)}(\xi)+\mathcal{S}_1^{(2)}(\xi)+\mathcal{S}_2^{(2)}(\xi)$.
Here, $\mathcal{J}^{(2)}(\xi)-\mathcal{J}^{(2)}(L/\ell-\xi)$ and $\mathcal{S}_i^{(2)}(\xi)-\mathcal{S}_i^{(2)}(L/\ell-\xi)$ correspond to $-\tau_0\partial j_{s;zz}^{(2)}/\partial z$ and $\beta_iE_i^{(2)}$, respectively, with a multiplicative factor $(q\mathcal{E}_0\tau_0)^2m\alpha/4\pi^2$.
The three contributions are given by
\begin{subequations}
 \begin{align}
 \mathcal{J}^{(2)}(\xi) &=
 \sqrt{\frac{8}{9\pi}}\,R_2(\xi)-3\tilde{F}_3(\xi)+\tilde{F}_5(\xi) -\Theta(\xi)[2\xi\tilde{F}_4(\xi)-\xi^2\tilde{F}_1(\xi)],\\
  \mathcal{S}_1^{(2)}(\xi) &= 2\tilde{F}_3(\xi)-2\tilde{F}_5(\xi),\\
  \mathcal{S}_2^{(2)}(\xi) &= -\sqrt{\frac{8}{9\pi}}\,R_2(\xi),
 \end{align}
\end{subequations}
where $\Theta(\xi)$ is the Heaviside step function.
Finally, we show the expression of the momentum flux tensor $\Pi_{zz}(z)$ in the limit of $\lambda_{\mathrm{TF}}/\ell \to 0$:
\begin{align}
    &\Pi_{zz}(\xi)/\left(\frac{m^2v_\mathrm{F}(q\mathcal{E}_0\ell)^2}{3\pi^2}\right)=\Pi_{zz}(\xi)/\left(\tau_0\mathcal{E}_0j_0\right)=-\int_{1}^{\infty} dx \frac{x^2[\Theta(\xi)-\mathrm{sgn}(\xi)e^{-x |\xi|}/2]}{(1 - x^2) \left\{ \left[x + \frac{1}{2} \ln\left(\frac{x-1}{x+1}\right)\right]^2 + \frac{\pi^2}{4} \right\}}.
\end{align}
%%%%%%%%%%%%%%%%%%%%%%%%%%%%%%%%%
\subsection{Spin distribution under other choices of the boundary condition for the current injection}
In the main text, we chose the simplest source term [Eq.~\eqref{eq: Jkgamma_1}].
\begin{figure*}
 \includegraphics[pagebox=artbox,width=0.95\textwidth]{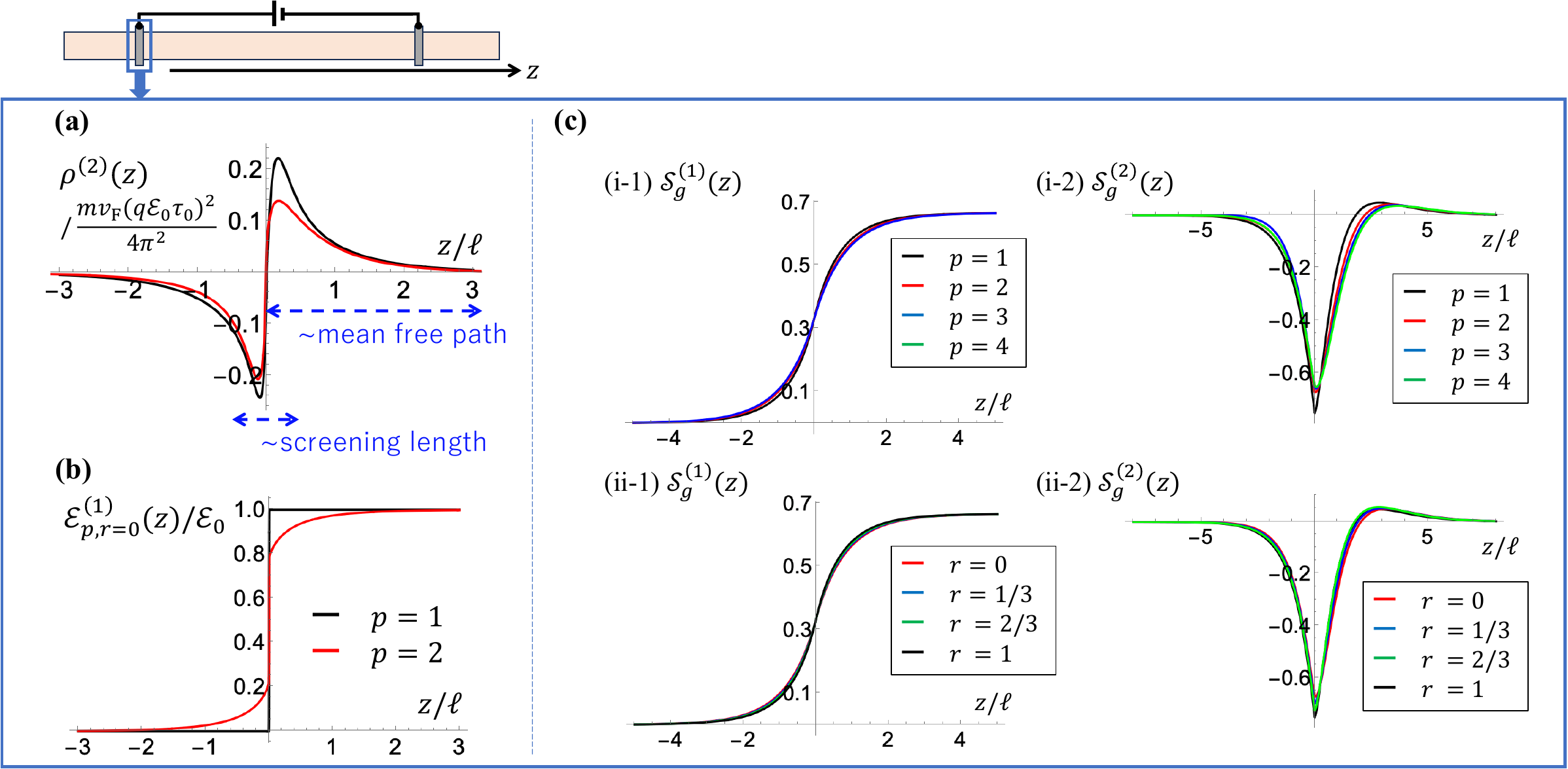}
 \caption{
 Enlarged view of spatial variations in electric quantities and the spin polarization near the interface for $q=-e$ and $\alpha>0$.
 (a) The excess charge density $\rho^{(2)}(z)$ shown for $\lambda_{\rm TF} / \ell = 0.1$.
 (b) The distributions of the electrochemical potential gradient of the order of $j_0$, $\mathcal{E}_{p,r}^{(1)}(z)$, are shown.
Ohm's law as a local relation does not hold near the surface, within a distance on the order of the mean free path $\ell$.
For clarity, the cases $p=1, r=0$ (black curve, where Ohm's law as a local relation holds throughout the system) and $p=2, r=0$ (red curve) are shown.
Results for other combinations of $\{p,r\}$ are qualitatively similar to those for $\{p=2,r=0\}$. (c) The spin polarization $s_z^{(1)}(z)$ and $s_z^{(2)}(z)$ under condition (i) $J_{\bm{k}\gamma}=J_{p,\bm{k}\gamma}$ for $p=1\text{--}4$ and (ii) $J_{\bm{k}\gamma}=rJ_{p=1,\bm{k}\gamma}+(1-r)J_{p=2,\bm{k}\gamma}$ for $r=0,1/3,2/3,1$.
 }
 \label{fig: fig4_test}
\end{figure*}
We here consider the spin distribution for more general forms of the source term,
\begin{align}
 J_{\bm{k}\gamma} &= r J_{1,\bm{k}\gamma} + (1-r) J_{p\ge2,\bm{k}\gamma},
 \label{eq: Jkgamma} \\
 J_{p,\bm{k}\gamma} &= -\frac{(2p+1)j_0}{2q N_\gamma}\left(\frac{v_{\bm{k}\gamma}^z}{\vF}\right)^{2p}\frac{\partial f^{(0)}(\varepsilon_{\bm{k}\gamma})}{\partial \varepsilon_{\bm{k}\gamma}},
\end{align}
with $p \in \mathbb{Z}_{>0}$ and $r \in [0, 1]$.
This form also ensures Eq.~\eqref{eq: source-sum-rule} and the continuity of spin density/current at $z = 0, \, L$ for general $(p, q)$.
When $r = 1$ ($J_{\bm{k}\gamma} = J_{1,\bm{k}\gamma}$), this form coincides with Eq.~\eqref{eq: Jkgamma_1}, and Ohm's law is satisfied throughout the system.
It should be noted that, in this context, Ohm's law relates the current to the gradient of the electrochemical potential rather than to the electric field.

In the following, we discuss two cases: (i) $p\ge2, \, r=0$ and (ii) $p=2, \, 0\le r\le1$.
In both cases, the electrochemical potential gradient, namely, the effective electric field, $\mathcal{E}_{p,r}^{(1)}(z)$ of the first order in $j_0$ is obtained as
\begin{align}
 \mathcal{E}_{p,r}^{(1)}(z)=\mathcal{E}_0\left[\Theta(z)-(1-r)\mathrm{sgn}(z)\eta_{p}(z)\right]
\end{align}
with
\begin{equation}
 \eta_p(z):=\frac{2p+1}{6}\int_1^{\infty}dx\frac{e^{-x|z|/l}\sum_{m=2}^{p}\frac{x^{-2(p-m)}}{2m-1}}{\left[x+\frac12\ln\left(\frac{x-1}{x+1}\right)\right]^2+\frac{\pi^2}{4}}.
\end{equation}
This results in the local violation of Ohm's law as shown in Fig.~\ref{fig: fig4_test}(b).
The spin polarizations $s_z^{(1)}(z)$ and $s_z^{(2)}(z)$ are obtained by the same method as in the main text, using the Boltzmann equation \eqref{eq: Boltzmann equation} with Eqs.~\eqref{eq: St-constant-W} and \eqref{eq: source term}, and Gauss's law.
The first-order spin polarization $s_z^{(1)}(z)$ is given by
\begin{equation}
 s_z^{(1)}(z)=-\frac{q\mathcal{E}_0\tau_0}{2\pi^2}m^2\alpha\vF\left[\mathcal{S}_g^{(1)}(z/l)-\mathcal{S}_g^{(1)}((L-z)/l)\right],
 \label{eq: s1_general}
\end{equation}
with
\begin{equation}
 \mathcal{S}_g^{(1)}(\xi):=\frac23\Theta(\xi)-\left[r\tilde{F}_4(\xi)+(1-r)\frac{2p+1}{3}\tilde{F}_{2p+2}(\xi)\right].
\end{equation}
%Since 
The exact expression of $s_z^{(2)}(z)$ in the quadratic response is more complicated. %
%, it is presented in the SM~\cite{Note1}.
In the limit of $\lambda_{\rm TF}/\ell \to 0$, however, the expression reduces to
\begin{equation}
 s_z^{(2)}(z)=\frac{(q\mathcal{E}_0\tau_0)^2}{4\pi^2}m\alpha\left[\mathcal{S}_g^{(2)}(z/l)-\mathcal{S}_g^{(2)}((L-z)/l)\right]
 \label{eq: s2_general}
\end{equation}
with
\begin{align}
    &\mathcal{S}_g^{(2)}(\xi):=\int_0^\xi d\xi'\frac{\mathcal{E}_{p,r}^{(1)}(\xi')}{\mathcal{E}_0}(\xi-\xi')\left\{r\left[3\tilde{F}_1(\xi)-\tilde{F}_3(\xi)\right]+(1-r)\frac{2p+1}{3}\left[3\tilde{F}_{2p-1}(\xi)-\tilde{F}_{2p+1}(\xi)\right]\right\}\notag\\
   &+\left(\int_{-\infty}^{\xi} d\xi'\int_{-\infty}^{\xi'} d\xi''-\int_{\xi}^{\infty}d\xi'\int_{\xi'}^{\infty}d\xi''\right)\frac{\mathcal{E}_{p,r}^{(1)}(\xi')}{\mathcal{E}_0}\frac{\mathcal{E}_{p,r}^{(1)}(\xi'')}{\mathcal{E}_0}\left\{2\tilde{F}_3(\xi-\xi'')+(\xi-\xi')\left[\tilde{F}_0(\xi-\xi'')-\tilde{F}_2(\xi-\xi'')\right]\right\}.
\end{align}

Figure~\ref{fig: fig4_test}(c) shows the results of Eqs.~\eqref{eq: s1_general} and \eqref{eq: s2_general} in the cases (i) and (ii).
This demonstrates that the spin polarization has a sign opposite to that expected from the spin current under DC current injection for more general forms of $J_{\bm{k}\gamma}$.
This feature is further supported by the fact that the characteristic dipole-like charge distribution is preserved even under the breakdown of local Ohm's law, as shown in Fig.~\ref{fig: fig4_test}(b).

%%%%%%%%%%%%%%%%%%%%%%%%%%%%%%%%%%%%%%%%%%%%%%%%%%%%%%%
\section{\label{appendix_DetailedCalculations}Detailed Calculations}

In Appendix B, we present the derivation of the linear and quadratic responses induced by a DC electric current in two steps.
In section 1(2), we summarize the derivation of linear(quadratic) responses.
In section 3, we summarize the variations in linear and quadratic responses resulting from changes in the boundary conditions.

As a prerequisite,  we introduce for later convenience the dimensionless length $w:=z/\ell$, the carrier density $\tilde{n} := \int d\varepsilon n(\varepsilon)$, and the effective electric field $\mathcal{E}$ as
\begin{equation}
\mathcal{E}:=E-\frac{1}{q N_0 }\frac{\partial \tilde{n}}{\partial z}
=E-\frac{1}{q N_0 \ell }\frac{\partial \tilde{n}}{\partial w}=E-\frac{\epsilon_0}{q^2  N_0 \ell^2 }\frac{\partial^2 E}{\partial w^2},
\label{eq: def-of-E-tilde}
\end{equation}
where we use the Gauss's law $\partial E/\partial z= q \tilde{n}/\epsilon_0$ in the last equality.
Using Eq.~\eqref{eq: def-of-E-tilde}, the electric field $E$ and the carrier density $\tilde{n}$ are given by
\begin{equation}
 E(w)= \frac{\ell}{2\lambda_{\rm TF}} \int_{-\infty}^\infty dw' e^{-\ell|w-w'|/\lambda_{\rm TF}}\mathcal{E}(w'), \quad
 \tilde{n}(w)= -\frac{\epsilon_0 \ell}{2q \lambda_{\rm TF}^2} \int_{-\infty}^\infty dw' \frac{(w-w')}{|w-w'|}e^{-\ell|w-w'|/\lambda_{\rm TF}}\mathcal{E}(w').
 \label{eq: E-and-n-formal-solution}
\end{equation}
The electric current density and spin density are also expressed in terms of $\mathcal{E}$ in simple forms as shown below. 
\subsection{Linear response}
First, we consider the linear response.
We expand the Boltzmann equation [Eq.~\eqref{eq: Boltzmann equation} in the main text] with respect to $j_0$; for the first order in $j_0$, we obtain
\begin{equation}
  v_{\bm{k}\gamma}^z\frac{\partial f_{\bm{k}\gamma}^{(1)}}{\partial z}-qE^{(1)}(z)\left(-\frac{\partial f_{\bm{k}\gamma}^{(0)}}{\partial k_z}\right)=-\frac{f_{\bm{k}\gamma}^{(1)}}{\tau_0}+\frac{n^{(1)}(\varepsilon_{\bm{k}\gamma},z)}{\tau_0N_0}+\frac{3j_0}{2qN_\gamma}\cos^2{\theta}\left(-\frac{\partial f_{\bm{k}\gamma}^{(0)}}{\partial \varepsilon_{\bm{k}\gamma}}\right)\delta(z).
  \label{eq_BE1_v2}
\end{equation}
The solution of Eq.~\eqref{eq_BE1_v2} that satisfies $f_{\bm{k}\gamma}^{(1)} \to 0$ for $z \to -\infty$ is given by %(Yoshimi (4.10))
\begin{equation}
  \begin{cases}
      \mathbf{\fbox{For $z<0$}}\quad&\\
      v_{\bm{k}\gamma}^z>0 : f_{\bm{k}\gamma}^{(1)}(z)=&\displaystyle\int_{-\infty}^{z}dz'\exp{\left(-\frac{z-z'}{v_{\bm{k}\gamma}^z\tau_0}\right)}\left(qE^{(1)}(z')+\frac{\tilde{n}^{(1)}(z')}{v_{\bm{k}\gamma}^z\tau_0N_0}\right)\left(-\frac{\partial f_{\bm{k}\gamma}^{(0)}}{\partial \varepsilon_{\bm{k}\gamma}}\right),\\
      v_{\bm{k}\gamma}^z<0 : f_{\bm{k}\gamma}^{(1)}(z)=&\displaystyle\int_{+\infty}^{z}dz'\exp{\left(-\frac{z-z'}{v_{\bm{k}\gamma}^z\tau_0}\right)}\left(qE^{(1)}(z')+\frac{\tilde{n}^{(1)}(z')}{v_{\bm{k}\gamma}^z\tau_0N_0}\right)\left(-\frac{\partial f_{\bm{k}\gamma}^{(0)}}{\partial \varepsilon_{\bm{k}\gamma}}\right)\\
      &\displaystyle+\frac{3j_0}{v(\varepsilon_{\bm{k}\gamma})\cdot 2qN_{\gamma}}|\cos{\theta}|\left(-\frac{\partial f_{\bm{k}\gamma}^{(0)}}{\partial \varepsilon_{\bm{k}\gamma}}\right)\cdot\exp{\left(-\frac{|z|}{|v_{\bm{k}\gamma}^z|\tau_0}\right)},\\
      \mathbf{\fbox{For $z>0$}}\quad&\\
      v_{\bm{k}\gamma}^z>0 : f_{\bm{k}\gamma}^{(1)}(z)=&\displaystyle\int_{-\infty}^{z}dz'\exp{\left(-\frac{z-z'}{v_{\bm{k}\gamma}^z\tau_0}\right)}\left(qE^{(1)}(z')+\frac{\tilde{n}^{(1)}(z')}{v_{\bm{k}\gamma}^z\tau_0N_0}\right)\left(-\frac{\partial f_{\bm{k}\gamma}^{(0)}}{\partial \varepsilon_{\bm{k}\gamma}}\right)\\
      &\displaystyle+\frac{3j_0}{v(\varepsilon_{\bm{k}\gamma})\cdot 2qN_{\gamma}}\cos{\theta}\left(-\frac{\partial f_{\bm{k}\gamma}^{(0)}}{\partial \varepsilon_{\bm{k}\gamma}}\right)\cdot\exp{\left(-\frac{z}{v_{\bm{k}\gamma}^z\tau_0}\right)},\\
      v_{\bm{k}\gamma}^z<0 : f_{\bm{k}\gamma}^{(1)}(z)=&\displaystyle\int_{+\infty}^{z}dz'\exp{\left(-\frac{z-z'}{v_{\bm{k}\gamma}^z\tau_0}\right)}\left(qE^{(1)}(z')+\frac{\tilde{n}^{(1)}(z')}{v_{\bm{k}\gamma}^z\tau_0N_0}\right)\left(-\frac{\partial f_{\bm{k}\gamma}^{(0)}}{\partial \varepsilon_{\bm{k}\gamma}}\right),
  \end{cases}
%  \label{eq_f1_cases}
\label{eq: f_(1)}
\end{equation}
where we use $n^{(1)}(\varepsilon_{\bm{k}\gamma}, z) \simeq \tilde{n}^{(1)}(z) \left(-\frac{\partial f_{\bm{k}\gamma}^{(0)}}{\partial \varepsilon_{\bm{k}\gamma}}\right)$, which is valid at low temperatures.
The velocity is defined by $v(\varepsilon_{\bm{k}\gamma}) := \frac{\partial \varepsilon_{\bm{k}\gamma}}{\partial k} = \sqrt{\alpha^2 + \frac{2\varepsilon_{\bm{k}\gamma}}{m}}$ for $\varepsilon_{\bm{k}\gamma} \geq 0$.
Substituting the solutions Eq.~\eqref{eq: f_(1)} into the definition of $\tilde{n}$, we obtain the integral equation for $\tilde{n}^{(1)}$ and $E^{(1)}$:
\begin{equation}
 \tilde{n}^{(1)}(w) =  \frac{3j_0}{2qv_{\mathrm{F}}} \tilde{F}_3(w) + \frac{q\ell N_0}{2} [\tilde{F}_2 * E^{(1)}](w) + \frac{1}{2} [\tilde{F}_1 * \tilde{n}^{(1)}](w),
 \label{eq: eq-for-n-tilde-1}
\end{equation}
where $\tilde{F}_n(w)$ is a special function defined by
\begin{gather}
 F_n(w) := \int_{0}^{1} dx x^{n-2} e^{-w/x} \quad (n \in \mathbb{Z}, \, w > 0), \qquad
 \tilde{F}_n(w) :=
 \begin{cases}
  F_n(|w|) & n = \text{odd}, \\
  \sgn(w) F_n(|w|) & n = \text{even},
 \end{cases}
\end{gather}
and $[A * B]$ represents a convolution:
\begin{equation}
 [A * B](w) := \int_{-\infty}^{\infty} dw' A(w - w') B(w').
\end{equation}
By integrating the third term in Eq.~\eqref{eq: eq-for-n-tilde-1} by parts, we obtain
\begin{equation}
 \frac{1}{2} [\tilde{F}_1 * \tilde{n}^{(1)}](w)
 = \tilde{n}^{(1)}(w) - \frac{1}{2} \left[\tilde{F}_2 * \frac{\partial \tilde{n}^{(1)}}{\partial w}\right](w).
\end{equation}
With the use of Eq.~\eqref{eq: def-of-E-tilde}, therefore, Eq.~\eqref{eq: eq-for-n-tilde-1} is rewritten in a compact form,
\begin{equation}
[ \tilde{F}_2*\mathcal{E}^{(1)}] (w)=-\frac{3 j_0 \tilde{F}_3(w)}{q^2 \vF N_0\ell }. % .(4.11) in mthesis 
\label{eq: eq-for-E-tilde-1}
\end{equation}
Furthermore, by integrating Eq.~\eqref{eq: eq-for-E-tilde-1} with respect $w$ from $-\infty$ to $w$ and using the relation 
\begin{equation}
  \int_{-\infty}^w dw' \tilde{F}_3(w')=\frac23\Theta(w)-\tilde{F}_4(w),
\end{equation}
we obtain the following equation:
\begin{equation}
    [ \tilde{F}_3*\mathcal{E}^{(1)}] (w)=-\frac{ j_0 }{q^2 \vF N_0\ell }\left(3\tilde{F}_4(w)-2\Theta(w)\right).
\label{eq: eq-for-E-tilde-1-integral}
\end{equation}

Next, we discuss the electric current density and the spin density.
Substituting the solutions Eq.~\eqref{eq: f_(1)} into the definition of them, we obtain 
\begin{align}
  &j_e^{(1)} (w)=\frac{3j_0 \tilde{F}_4(w)}{2}+\frac{q^2 v_{\rm F}N_0 \ell}{2}[\tilde{F}_3*\mathcal{E}^{(1)} ](w), \\% mthesis p.31 sec.4.3
  &s_z^{(1)} (w)=-\frac{m^2 \alpha \ell q}{2\pi^2}[\tilde{F}_3*\mathcal{E}^{(1)} ](w).% mthesis p.33 sec. 4.4
  \label{eq: sz1_with_E1}
  \end{align}
We can derive, without solving Eq.~\eqref{eq: eq-for-E-tilde-1}, the expressions for $j^{(1)}(w)$ and $s_z^{(1)}(w)$ as
\begin{align}
   & j_e^{(1)} (w)=j_0 \Theta(w)\label{eq: exp-for-j1}, \\
   & s_z^{(1)} (w)=-\frac{m^2 \alpha  j_0}{2\pi^2q \vF N_0}\left(2\Theta(w)-3\tilde{F}_4(w)\right), % p.34 in mthesis 
\label{eq: exp-for-s1}
\end{align}
by using the relation \eqref{eq: eq-for-E-tilde-1-integral}.

Finally, we derive the electric field and the carrier density.
We can find the solution to Eq.~\eqref{eq: eq-for-E-tilde-1} as
\begin{align}
 \mathcal{E}^{(1)}(w)=\frac{3 j_0 \Theta(w)}{q^2 \vF N_0\ell }=:\mathcal{E}_0 \Theta(w),
 \label{eq: E-tilde-1-final}
\end{align}
with use of the relation 
\begin{align}
 \int_{-\infty}^w \tilde{F}_2(w')dw'=-\tilde{F}_3(w).
\end{align}
By using Eqs.~\eqref{eq: E-and-n-formal-solution} and \eqref{eq: E-tilde-1-final}, therefore, we obtain the explicit expressions for $E$ and $\tilde{n}$: % p.33 sec. 4.2 in mthesis
\begin{align}
&E^{(1)}(w)=\mathcal{E}_0 \left(\Theta(w)-\frac{w e^{-\ell|w|/\lambda_{\rm TF}}}{2|w|}\right),
\label{eq: E-1-final}\\
&\tilde{n}^{(1)}(w)=\frac{\epsilon_0\mathcal{E}_0}{2q\lambda_{\rm TF}} e^{-\ell|w|/\lambda_{\rm TF}}.
\label{eq: n-1-final}
\end{align}
By substituting these equations into Eq.~\eqref{eq: f_(1)} and integrating it by parts, the distribution function of $O(j_0)$ is given by
\begin{equation}
 \left\{
 \begin{aligned}
  & \mathbf{\fbox{For $z < 0$}} \\
  &
  \begin{aligned}
   v_{\bm{k}\gamma}^z > 0 : f_{\bm{k}\gamma}^{(1)}(z) &=
   \frac{\tilde{n}^{(1)}(z)}{N_0}
   \left(-\frac{\partial f_{\bm{k}\gamma}^{(0)}}{\partial \varepsilon_{\bm{k}\gamma}}\right), \\
   v_{\bm{k}\gamma}^z<0 : f_{\bm{k}\gamma}^{(1)}(z) &=
   \left[ \frac{\tilde{n}^{(1)}(z)}{N_0} + h_\gamma(\varepsilon_{\bm{k}\gamma}, \cos\theta, z) \right]
   \left(-\frac{\partial f_{\bm{k}\gamma}^{(0)}}{\partial \varepsilon_{\bm{k}\gamma}}\right),
  \end{aligned} \\
  & \mathbf{\fbox{For $z > 0$}} \\
  &
  \begin{aligned}
   v_{\bm{k}\gamma}^z>0 : f_{\bm{k}\gamma}^{(1)}(z) &=
   \left[ \frac{\tilde{n}^{(1)}(z)}{N_0} + h_\gamma(\varepsilon_{\bm{k}\gamma}, \cos\theta, z) +q \mathcal{E}_0 v(\varepsilon_{\bm{k}\gamma}) \tau_0 \cos\theta \right]
   \left(-\frac{\partial f_{\bm{k}\gamma}^{(0)}}{\partial \varepsilon_{\bm{k}\gamma}}\right), \\
   v_{\bm{k}\gamma}^z<0 : f_{\bm{k}\gamma}^{(1)}(z) &=
   \left[ \frac{\tilde{n}^{(1)}(z)}{N_0} -q\mathcal{E}_0 v(\varepsilon_{\bm{k}\gamma}) \tau_0 \lvert\cos\theta\rvert \right]
   \left(-\frac{\partial f_{\bm{k}\gamma}^{(0)}}{\partial \varepsilon_{\bm{k}\gamma}}\right),
  \end{aligned}
 \end{aligned}
 \right.
 \label{eq: f_(1)-final}
\end{equation}
where we define
\begin{equation}
 h_\gamma(\varepsilon_{\bm{k}\gamma}, \cos\theta, z)
 := \left(-q \mathcal{E}_0 v(\varepsilon_{\bm{k}\gamma}) \tau_0 + \frac{3j_0}{v(\varepsilon_{\bm{k}\gamma}) \cdot 2q N_\gamma} \right)
 \lvert\cos\theta\rvert \exp\left(- \frac{\lvert z \rvert}{v(\varepsilon_{\bm{k}\gamma}) \tau_0 \lvert\cos\theta\rvert}\right).
\end{equation}

\subsection{Quadratic response}
Next, we discuss the Boltzmann equation of $O((j_0)^2)$:% mthesis (5.1) in p.39
\begin{equation}
  \left(\frac{\partial f_{\bm{k}\gamma}^{(2)}}{\partial t}+\right)v_{\bm{k}\gamma}^z\frac{\partial f_{\bm{k}\gamma}^{(2)}}{\partial z}-qE^{(2)}(z) v_{\bm{k}\gamma}^z \left(-\frac{\partial f_{\bm{k}\gamma}^{(0)}}{\partial\varepsilon_{\bm{k}\gamma}}\right)-qE^{(1)}(z)\left(-\frac{\partial f_{\bm{k}\gamma}^{(1)}}{\partial k_z}\right)=-\frac{f_{\bm{k}\gamma}^{(2)}}{\tau_0}+\frac{n^{(2)}(\varepsilon_{\bm{k}\gamma},z)}{\tau_0 N_0}.
  \label{eq_BE2}
\end{equation}
In contrast to the linear response, the Boltzmann equation with only elastic scattering has no stationary solution due to Joule heat, which originates from the fourth term in the LHS of Eq.~\eqref{eq_BE2}.
Thus, in the following, $n^{(2)}(\varepsilon,z)$ is approximated to $n^{(2)}(\varepsilon,z)\sim\tilde{n}^{(2)}(z)\delta(\varepsilon-\mu)$ so that $f_{\bm{k}\gamma}^{(2)}(z)=-qE^{(1)}(z)\tau_0(\partial f_{\bm{k}\gamma}^{(1)}/\partial k_z)$ is satisfied in the bulk region and the whole system is in a steady state.
This approximation can also be interpreted as adding $-[n^{(2)}(\varepsilon,z)-\tilde{n}^{(2)}(z)\delta(\varepsilon-\mu)]/\tau_0 N_0$, which corresponds to inelastic scatterings that dissipate Joule heat, to the collision term.
We then set this contribution to $-f_{\bm{k}\gamma}^{(2)}/\tau_{\mathrm{inel}}$ and evaluate relaxation time $\tau_{\mathrm{inel}}$.
By multiplying $\varepsilon_{\bm{k}\gamma}/\Omega$ to Eq.~\eqref{eq_BE2} and summing over $\bm{k}$ and $\gamma$, we obtain the energy transport equation,
\begin{equation}
    \frac{\partial\mathcal{E}^{(2)}_u}{\partial t}+\frac{\partial j^{(2)}_u}{\partial z}+qE^{(1)}(z)\frac{1}{\Omega}\sum_{\bm{k}\gamma}\varepsilon_{\bm{k}\gamma}\frac{\partial f_{\bm{k}\gamma}^{(1)}}{\partial k_z}=-\frac{\mathcal{E}^{(2)}_u}{\tau_{\mathrm{inel}}}
    \label{eq: energy_transport_equation}
\end{equation}
with $\mathcal{E}^{(2)}_u(t,z):=\sum_{\bm{k}\gamma}\varepsilon_{\bm{k}\gamma}f^{(2)}_{\bm{k}\gamma}/\Omega$, and $j^{(2)}_u(t,z):=\sum_{\bm{k}\gamma}\varepsilon_{\bm{k}\gamma}v^z_{\bm{k}\gamma}f^{(2)}_{\bm{k}\gamma}/\Omega$.
In the steady state, $\partial \mathcal{E}^{(2)}_u / \partial t \rightarrow0$. 
In the bulk, $\partial j_u^{(2)} / \partial z \rightarrow0$, and 
$f_{\bm{k}\gamma}^{(2)}(z) = - q E^{(1)}(z) \tau_0 (\partial f_{\bm{k}\gamma}^{(1)} / \partial k_z)$ holds. 
Under these conditions, the energy transport equation~\eqref{eq: energy_transport_equation} gives $\tau_{\mathrm{inel}} = \tau_0$ for the energy relaxation time.
Temperature change due to the Joule heating  is negligible in a steady state under a condition  \color{black}    $T\gg\frac{\mathrm{\tau_{\mathrm{inel}}\cdot[Joule~Heat]}}{\mathrm{specific} \ \mathrm{heat}}=\frac{3\tau_0 j_0^2}{\sigma_0 k_\mathrm{B}^2 T m k_\mathrm{F}}.$
Using the carrier density $n_0=\frac{1}{\Omega}\sum_{\bm{k}\gamma}f^{(0)}_{\bm{k}\gamma}$ in equilibrium, this condition reduces to
\begin{equation}
    \frac{j_0}{qn_0 v_\mathrm{F}}\ll\frac{k_\mathrm{B}T}{\mu},
\end{equation}
which can be satisfied in actual metals, where  $|j_0/qn_0 v_\mathrm{F}|\simeq e\mathcal{E}_o\tau_0/k_\mathrm{F}$ holds. \color{black}
Lastly, we emphasize that this approximation gives $\tilde{n}^{(2)}(z)/N_0=\sum_{\bm{k}\gamma}f_{\bm{k}\gamma}^{(2)}(z)/\Omega$ and thus does not affect the charge conservation law.

The solution of Eq.~\eqref{eq_BE2} that satisfies $f_{\bm{k}\gamma}^{(2)} \to 0$ for $z \to \pm\infty$ is given by % Yoshimi mthesis (5.2) in p.41
\begin{equation}
 \left\{
 \begin{aligned}
  & \underline{v_{\bm{k}\gamma}^z > 0 : } \\
  & f_{\bm{k}\gamma}^{(2)}(z)
  = \displaystyle\int_{-\infty}^{z}dz'\exp{\left(-\frac{z-z'}{v_{\bm{k}\gamma}^z\tau_0}\right)}\left[qE^{(2)}(z')\left(-\frac{\partial f_{\bm{k}\gamma}^{(0)}}{\partial \varepsilon_{\bm{k}\gamma}}\right)+\frac{n^{(2)}(\varepsilon_{\bm{k}\gamma}, z')}{v_{\bm{k}\gamma}^z\tau_0 N_0}+\frac{qE^{(1)}(z')}{v_{\bm{k}\gamma}^z}\left(-\frac{\partial f_{\bm{k}\gamma}^{(1)}}{\partial k_z}\right)\right], \\
  & \underline{v_{\bm{k}\gamma}^z < 0 : } \\
  & f_{\bm{k}\gamma}^{(2)}(z)
  = \displaystyle\int_{+\infty}^{z}dz'\exp{\left(-\frac{z-z'}{v_{\bm{k}\gamma}^z\tau_0}\right)}\left[qE^{(2)}(z')\left(-\frac{\partial f_{\bm{k}\gamma}^{(0)}}{\partial \varepsilon_{\bm{k}\gamma}}\right)+\frac{n^{(2)}(\varepsilon_{\bm{k}\gamma}, z')}{v_{\bm{k}\gamma}^z\tau_0 N_0}+\frac{qE^{(1)}(z')}{v_{\bm{k}\gamma}^z}\left(-\frac{\partial f_{\bm{k}\gamma}^{(1)}}{\partial k_z}\right)\right].
 \end{aligned}
 \right.
 \label{eq: f_(2)}
\end{equation}
Substituting these solutions Eq.~\eqref{eq: f_(2)} into the definition of $\tilde{n}$, and using Eq.~\eqref{eq: f_(1)-final}, we obtain % ( 5.5) in P. 42 in mthesis
\begin{equation}
 \tilde{n}^{(2)}(w) = \frac{q\ell N_0}{2} [\tilde{F}_2 * E^{(2)}](w) + \frac{1}{2} [\tilde{F}_1 * \tilde{n}^{(2)}](w) - \frac{q\ell N_0}{2} X(w),
 \label{eq: eq-for-n-tilde-2}
\end{equation}
with % ( 5.4) in P. 42 in mthesis
\begin{align}% Mthesis (5.4)
  X(w)=&-\frac{2\mu }{\pi^2v_F^3N_0^2}\int_{-\infty}^{+\infty}dw'E^{(1)}(w')\tilde{n}^{(1)}(w')\left[\tilde{F}_2\left(w-w'\right)+(w-w')\tilde{F}_1\left(w-w'\right)\right]\notag\\
  &+\frac{3 m j_0}{\pi^2 q N_0^2 v_\mathrm{F}^2 }\int_0^{+\infty}dw'E^{(1)}(w')\left[\tilde{F}_1\left(w-w'\right)-(w-w')\tilde{F}_0\left(w-w'\right)\right],
  \label{eq_**_solved}
\end{align}
where we neglect contributions from the second and higher orders in $\alpha$.
(We apply the same treatment to the following calculations.)
Using integration by parts, Eq.~\eqref{eq: eq-for-n-tilde-2} is rewritten in a compact form,
\begin{equation}
[ \tilde{F}_2*\mathcal{E}^{(2)}] (w) = X(w).
\label{eq: eq-for-E-tilde-2}
\end{equation}

Next, we discuss the electric current density and the spin density.
Substituting the solutions Eq.~\eqref{eq: f_(2)} into their definitions, we obtain 
\begin{align}
  &j_e^{(2)} (w)=\frac{q^2 v_{\rm F}N_0 \ell}{2}[\tilde{F}_3*\mathcal{E}^{(2)} ](w)+X_j(w)\label{eq: j2}, \\% (5.19) in P. 49 mthesis
  &s_z^{(2)} (w)=-\frac{m^2 \alpha \ell q}{2\pi^2}[\tilde{F}_3*\mathcal{E}^{(2)} ](w)+X_s(w),
   \label{eq: s2}% sec 5.4.1 in p.50 mthesis
\end{align}
with 
\begin{subequations}    
\begin{align}
  X_j(w)&=\frac{q^2 m\ell}{2\pi^2N_0}\int_{-\infty}^{+\infty}dw'E^{(1)}(w')\tilde{n}^{(1)}(w')\left[2\tilde{F}_3\left(w-w'\right)+\tilde{F}_2\left(w-w'\right)(w-w')\right]\label{eq: Xj-1}\\ %(4.17) in p.48 in mthesis
  &+\frac{3 q m \tau_0 j_0}{2\pi^2 N_0}\int_{0}^{+\infty}dw'E^{(1)}(w')\tilde{F}_{1}\left(w-w'\right)(w-w'),
  \label{eq: Xj-2}
\end{align}
\end{subequations}
and
\begin{subequations}
\begin{align}
X_s(w)
=&-\frac{mq\alpha \ell}{2\pi^2v_\mathrm{F}^2N_0}\int_{-\infty}^{+\infty}dw'E^{(1)}(w')\tilde{n}^{(1)}(w')\tilde{F}_2\left(w-w'\right)(w-w')\label{eq: Xs-1}\\
+&\frac{3\alpha m\tau_0 j_0}{4\pi^2 v_\mathrm{F}^2 N_0 }\left[-2 \tilde{F}_4(w)\int_{0}^{w}dw'E^{(1)}(w')+2\tilde{F}_1(w) \int_0^{w}dw'E^{(1)}(w')(w-w')\right]\label{eq: Xs-2}\\
 +&\frac{3\alpha m\tau_0 j_0}{4\pi^2 v_\mathrm{F}^2 N_0 }\int_{0}^{+\infty}dw'E^{(1)}(w')\left\{2\tilde{F}_4(w-w')-(w-w')\left[F_1(|w-w'|)+F_3(|w-w'|)\right]\right\}.
 \label{eq: Xs-3}
\end{align}
\end{subequations}
We can confirm $ j_e^{(2)}(w)=0$ by integrating Eq.~\eqref{eq: eq-for-E-tilde-2} with respect $w$ from $-\infty$ to $w$ and by using the relation
\begin{equation}
\int_{-\infty}^w X(w')dw'=\frac{2X_j(w) }{q^2 v_\mathrm{F}^2 \tau_0 N_0} ,    % p.49 mthesis
\end{equation}
which follows from 
\begin{equation}
\frac{d \tilde{F}_3(w)}{d w}=-\tilde{F}_2(w),
\quad 
\frac{d (w \tilde{F}_2(w))}{d w}=\tilde{F}_2(w)-w \tilde{F}_1(w),
\quad\frac{d (w \tilde{F}_1(w))}{d w}=\tilde{F}_1(w)-w \tilde{F}_0(w)
. 
\end{equation}
From Eq.~\eqref{eq: j2} ($= 0$),  and \eqref{eq: s2},  we find that 
\begin{equation}
 s_z^{(2)} (w)= \frac{m^2 \alpha  }{\pi^2 N_0 q v_\mathrm{F}}X_j(w)+X_s(w).
   \label{eq: s2-final}
\end{equation}
For $|w|\gg \lambda_{\rm TF}/\ell$,  the contribution from \eqref{eq: Xj-1} is smaller by $O(\lambda_{\rm TF}/\ell)$ than \eqref{eq: Xj-2} because the integrand in the former is localized near   $|w'|\sim \lambda_{\rm TF}/\ell$ while that in the latter is extended. Further, we replace $E^{(1)}(w')$ by $\mathcal{E}_0\Theta(w')$ in  \eqref{eq: Xj-2}  when we extract the  dominant contribution for $|w|\gg \lambda_{\rm TF}/\ell$. We then obtain 
\begin{equation}
 X_j \simeq \frac{3 q m \tau_0 j_0\mathcal{E}_0}{2\pi^2 N_0}\int_{0}^{+\infty}dw'\tilde{F}_{1}\left(w-w'\right)(w-w') = - \frac{3 q m \tau_0 j_0\mathcal{E}_0}{2\pi^2 N_0}\left[\tilde{F}_3(w)+w\tilde{F}_2(w)\right].
 \label{eq: Xj-simplified}
\end{equation}

Similarly, the dominant contribution in $X_s$ for $|w|\gg \lambda_{\rm TF}/\ell$,  stems from  \eqref{eq: Xs-2} and \eqref{eq: Xs-3}. Replacing  $E^{(1)}$ by $\mathcal{E}_0\Theta(w')$ in these terms, we then obtain 
\begin{equation}
 X_s \simeq   
 \frac{3\alpha m\tau_0 j_0\mathcal{E}_0}{4\pi^2 v_\mathrm{F}^2 N_0 }
\left\{\Theta(w)\left[-2w \tilde{F}_4(w)+w^2 \tilde{F}_1(w)\right]+w \left[\tilde{F}_2(w)+\tilde{F}_4(w)\right]+\tilde{F}_3(w)-\tilde{F}_5(w)\right\}.
\label{eq: Xs-simplified}
\end{equation}
From Eqs.~\eqref{eq: s2-final}, \eqref{eq: Xj-simplified}, 
 and  \eqref{eq: Xs-simplified},  we find that 
 \begin{equation}
 s_z^{(2)} (w)=  \frac{3\alpha m\tau_0 j_0\mathcal{E}_0}{4\pi^2 v_\mathrm{F}^2 N_0 }
\left\{\Theta(w)\left[-2w \tilde{F}_4(w)+w^2 \tilde{F}_1(w)\right]+w \left[\tilde{F}_4(w)-\tilde{F}_2(w)\right] -\tilde{F}_3(w)-\tilde{F}_5(w)\right\}.
\end{equation}

\subsection{Robustness against choices of the boundary conditions for the current injection}
Here, we consider more general boundary conditions to examine the robustness of the sign discrepancy against boundary variations.
To satisfy Eq.~\eqref{eq: source-sum-rule} and the continuity of spin and spin current, the possible form of $J_{\bm{k}\gamma}$ reduces to
\begin{align}
    J_{\bm{k}\gamma}=\sum_{p\in P}C_pJ_{p,\bm{k}\gamma}
\end{align}
with $P\subset\mathbb{Z}_{>0},\,C_p=\mathrm{Const.}$ and
\begin{align}
    \sum_{p\in P}C_p=1,\quad J_{p,\bm{k}\gamma}=-\frac{(2p+1)j_0}{2qN_{\gamma}}\left(\frac{v_{\bm{k}\gamma}^z}{v_\mathrm{F}}\right)^{2p}\frac{\partial f_{\bm{k}\gamma}^{(0)}}{\partial \varepsilon_{\bm{k}\gamma}}.
\end{align}
For simplicity, we treat the specific case $J_{\bm{k}\gamma}=rJ_{1,\bm{k}\gamma}+(1-r)J_{p,\bm{k}\gamma}$ with $p\ge2,\,0\le r\le1$.
Most of the calculations for the spin polarization in the linear and quadratic responses are the same as those in Sections 1 and 2.
Therefore, in this section, we focus on the distinct aspect—the form of the effective electric field.
By solving the Boltzmann equation in the same way as deriving Eq.~\eqref{eq: eq-for-E-tilde-1}, the effective electric field $\mathcal{E}_{p,r}^{(1)}(w)$ is expressed as
\begin{equation}
    \left[\tilde{F}_2*\mathcal{E}_{p,r}^{(1)}\right](w)=-\frac{3j_0}{q^2v_\mathrm{F}N_0\ell}\left[r\tilde{F}_3(w)+(1-r)\frac{2p+1}{3}\tilde{F}_{2p+1}(w)\right]
    \label{eq: CONV_F2_E_general}
\end{equation}
Since the equation
\begin{align}
    \int_{-\infty}^{w}\tilde{F}_{2p+1}(w')dw'=\frac{2}{2p+1}\Theta(w)-\tilde{F}_{2p+2}(w)
\end{align}
holds, Equation~\eqref{eq: CONV_F2_E_general} reduces to
\begin{equation}
    \left[\tilde{F}_3*\mathcal{E}_{p,r}^{(1)}\right](w)=-\frac{3j_0}{q^2v_\mathrm{F}N_0\ell}\left[r\tilde{F}_4(w)+(1-r)\frac{2p+1}{3}\tilde{F}_{2p+2}(w)\right]+\frac{2j_0}{q^2v_\mathrm{F}N_0\ell}\Theta(w).
\end{equation}
By applying this equation to Eq.~\eqref{eq: sz1_with_E1}, the spin polarization near the interface $z=0$ in the linear response is expressed as
\begin{align}
    s_z^{(1)}(w)=-\frac{q\mathcal{E}_0\tau_0}{2\pi^2}m^2\alpha v_\mathrm{F}\left\{2\Theta(w)-3\left[r\tilde{F}_4(w)+(1-r)\frac{2p+1}{3}\tilde{F}_{2p+2}(w)\right]\right\}.
\end{align}

Before calculating the spin polarization in the quadratic response, it is necessary to solve Eq.~\eqref{eq: CONV_F2_E_general}.
The scheme to do this is shown below:
(i) Transform Eq.~\eqref{eq: CONV_F2_E_general} into
\begin{equation}
    -\left[\tilde{F}_3*\mathcal{Y}_{p,r}^{(1)}\right](w)=-\frac{3j_0}{q^2v_\mathrm{F}N_0\ell}\left[r\tilde{F}_3(w)+(1-r)\frac{2p+1}{3}\tilde{F}_{2p+1}(w)\right]
    \label{eq: CONV_F3_Y}
\end{equation}
with $\mathcal{Y}_{p,r}^{(1)}(w):=\partial_w\mathcal{E}_{p,r}^{(1)}(w)$.
(ii) Solve Eq.~\eqref{eq: CONV_F3_Y} by using the Fourier transformation.
(iii) Calculate $\mathcal{E}_{p,r}^{(1)}(w)=\int_{-\infty}^wdw'\mathcal{Y}_{p,r}^{(1)}(w')$.
As a result, $\mathcal{E}_{p,r}^{(1)}(w)$ reduces to
\begin{align}
    \mathcal{E}_{p,r}^{(1)}(w)/\mathcal{E}_0=&r\,\Theta(w)\notag\\
    +&(1-r)\left\{\frac{2p+1}{3(2p-1)}\Theta(w)+\frac{2p+1}{6}\int_1^{+\infty}dx\frac{\sum_{m=2}^p\frac{1}{2m-1}x^{-2(p-m)}}{\left[x+\frac12\ln\left(\frac{x-1}{x+1}\right)\right]^2+\frac{\pi^2}{4}}\left[2\Theta(w)-\frac{w}{|w|}e^{-x|w|}\right]\right\}
\end{align}
near the interface $w=0$. In the limit $w \to +\infty$, $\mathcal{E}_{p,r}^{(1)}(w)$ converges to $\mathcal{E}_0$ due to
\begin{align}
    \int_1^{+\infty}dx\frac{\sum_{m=2}^p\frac{1}{2m-1}x^{-2(p-m)}}{\left[x+\frac12\ln\left(\frac{x-1}{x+1}\right)\right]^2+\frac{\pi^2}{4}}=\frac{4(p-1)}{4p^2-1}.
\end{align}
This indicates that local Ohm's law between the current density and effective electric field, breaks down near the interface within a range on the order of the mean free path $\ell$. 
\twocolumngrid
\bibliography{reference}
\end{document}